\documentclass[DIV=11]{scrartcl}
\usepackage[colorlinks=true,linkcolor=blue,urlcolor=blue,citecolor=blue,anchorcolor=green,pdfusetitle]{hyperref}

\usepackage[utf8]{inputenc}
\usepackage{microtype,dsfont,array,booktabs,makecell,multirow,siunitx}
\newcolumntype{C}[1]{>{\centering\arraybackslash$}m{#1}<{$}}
\usepackage[labelfont=it, textfont=it]{caption}

\usepackage{amsmath,amssymb}

\usepackage{xspace}
\input{braket}

\usepackage{tikz}
\usepackage[outline]{contour}
\contourlength{1.5pt}

\newcommand{\Liquid}{\textup{LIQ}\textit{Ui}$\ket{}$\xspace}
\newcommand{\Learn}{\Liquid-\textup{Learn}\xspace}

\newcommand{\C} {\mathds C}
\renewcommand{\H} {{\ensuremath{\mathcal H}}\xspace}

\newcommand{\1} {\ensuremath{\mathds 1}}
\newcommand{\ii}{\mathrm{i}}

\newcommand{\op}[1] {\mathbf{#1}}
\newcommand{\set}[1] {\mathrm{#1}}
\newcommand{\field}[1] {\mathds{#1}}
\newcommand{\PP}{\textnormal{\textsf{P}}\xspace}
\newcommand{\NP}{\textnormal{\textsf{NP}}\xspace}
\newcommand{\StoqMA}{\textnormal{\textsf{StoqMA}}\xspace}

\newcommand{\BQEXP}{\textnormal{\textsf{BQEXP}}\xspace}
\newcommand{\QMA}{\textnormal{\textsf{QMA}}\xspace}

\newcommand{\yes}{{\textnormal{\textsf{YES}}}\xspace}
\newcommand{\no}{{\textnormal{\textsf{NO}}}\xspace}

\newcommand{\LL}{\ensuremath{\set L}\xspace}
\newcommand{\Lyes}{\ensuremath{\set L_\textsf{YES}}\xspace}
\newcommand{\Lno}{\ensuremath{\set L_\textsf{NO}}\xspace}

\DeclareMathOperator{\diag}{diag}
\DeclareMathOperator{\supp}{supp}
\DeclareMathOperator{\tr}{tr}
\DeclareMathOperator{\gs}{\mathcal L}

\newtheorem{theorem}{Theorem}
\newtheorem{definition}[theorem]{Definition}

\usepackage{cleveref}
\crefname{lemma}{lemma}{lemmas}
\crefname{proposition}{proposition}{propositions}
\crefname{definition}{definition}{definitions}
\crefname{theorem}{theorem}{theorems}
\crefname{conjecture}{conjecture}{conjectures}
\crefname{corollary}{corollary}{corollaries}
\crefname{example}{example}{examples}
\crefname{section}{section}{sections}
\crefname{appendix}{appendix}{appendices}
\crefname{figure}{fig.}{figs.}
\crefname{equation}{eq.}{eqs.}
\crefname{table}{table}{tables}
\crefname{item}{property}{properties}
\crefname{remark}{remark}{remarks}

\usepackage[backend=biber,style=numeric-comp,doi=false,isbn=false,url=false,maxbibnames=20,sorting=none]{biblatex}
\bibliography{Bibliography/books}
\bibliography{Bibliography/neural}
\bibliography{Bibliography/hamilton}
\bibliography{Bibliography/supremacy}
\bibliography{Bibliography/commuting}
\newbibmacro{string+doi}[1]{\iffieldundef{doi}{#1}{\href{https://dx.doi.org/\thefield{doi}}{#1}}}
\DeclareFieldFormat{title}{\usebibmacro{string+doi}{\mkbibemph{#1}}}
\DeclareFieldFormat[article,thesis,incollection,inproceedings]{title}{\usebibmacro{string+doi}{\mkbibquote{#1}}}
\makeatletter
\DeclareFieldFormat{eprint:arxiv}{%
    \iffieldundef{eprintclass}{%
        \href{http://arxiv.org/\abx@arxivpath/#1}{arXiv\addcolon\nolinkurl{#1}}%
    }{%
        \href{http://arxiv.org/\abx@arxivpath/\thefield{eprintclass}/#1}{arXiv\addcolon\nolinkurl{#1}}%
        \addspace\UrlFont{\mkbibbrackets{\thefield{eprintclass}}}%
    }%
}
\DeclareFieldAlias{eprint:arXiv}{eprint:arxiv}
\makeatother
\renewbibmacro{in:}{}

\begin{document}
\title{Classifying Data with\\Local Hamiltonians}
\author{Johannes Bausch\thanks{DAMTP, University of Cambridge, Wilberforce Road, Cambridge CB3 0WA, UK---\texttt{jkrb2@cam.ac.uk}}}

\maketitle

\begin{abstract}
The goal of this work is to define a notion of a ``quantum neural network'' to classify data, which exploits the low energy spectrum of a local Hamiltonian.
As a concrete application, we build a binary classifier, train it on some actual data and then test its performance on a simple classification  task.
More specifically, we use Microsoft's quantum simulator, \Liquid, to construct local Hamiltonians that can encode trained classifier functions in their ground space, and which can be probed by measuring the overlap with test states corresponding to the data to be classified.
To obtain such a classifier Hamiltonian, we further propose a training scheme based on quantum annealing which is completely closed-off to the environment and which does not depend on external measurements until the very end, avoiding unnecessary decoherence during the annealing procedure.
For a network of size $n$, the trained network can be stored as a list of $O(n)$ coupling strengths.
We address the question of which interactions are most suitable for a given classification task, and develop a qubit-saving optimization for the training procedure on a simulated annealing device. Furthermore, a small neural network to classify colors into red vs. blue is trained and tested, and benchmarked against the annealing parameters.
\end{abstract}

\section{Introduction}
\subsection{Background and Context}
The field of Hamiltonian complexity theory has seen a lot of interesting contributions over the past decade. Based on Feynman's idea of embedding computation into a local Hamiltonian \cite{Feynman1986} and starting with Kitaev's original proof of \QMA-hardness of the ground state energy problem of 5-local Hamiltonians, the physical relevance of hardness results has been improved successively \cite{Kempe2006,Oliveira2008,Aharonov2009,gottesman2009quantum}.
More recent results include using history state and tiling Hamiltonian embeddings to prove undecidability of the spectral gap \cite{Cubitt2015} and size-driven phase transitions \cite{Bausch2015}, both of which embed highly complex computation into relatively simple local couplings.
Hamiltonians constructed from local interactions have been classified into the complexity classes \PP, \StoqMA, \NP and \QMA in \cite{Cubitt2013,Piddock2015}, and it is a well-known fact that almost all Hamiltonian interactions are universal in the sense that they can be used to approximate any 2-qubit unitary operation \cite{Childs2010,Cubitt2017}.
This suggests that local Hamiltonians are well-suited for computational tasks; but which concrete problems can they answer, and what is their implicit overhead?

Let us step back for a minute and assess what potential candidates we have for encoding computation into the low-energy spectrum of a Hamiltonian.
Roughly speaking, there are two fundamentally different embedding constructions available to date.

A so-called \emph{History State} construction is a local Hamiltonian allowing state transitions, based on Feynman's idea of embedding quantum computation into the ground state of the Hamiltonian. Given some Hilbert space \H and two states $\ket a,\ket b\in\H$, we can encode a transition rule (which represents e.g.\ a computational step) $a\leftrightarrow b$ by a local interaction of the form $(\ket a-\ket b)(\bra a-\bra b)$. The ground state of such a Hamiltonian will be a superposition of all states connected by transition rules. The most extensive such construction to date can be found in \cite{Bausch2016}, and the spectral gap of these History State Hamiltonians is known to close as $\Theta(1/T^2)$ in the system size, where $T$ is the number of computational steps taken \cite{Bausch2016a,Caha2017}.
The answer of the computation can be read out by measuring the ground state energy of the Hamiltonian; is it below a threshold $\alpha$, the outcome was \yes with high probability, if it is above $\beta$ it was likely \no; and partly as a result of the closing spectral gap the only promise we have on the two thresholds is $\beta-\alpha=\Omega(1/T^2)$, so extremely high precision is required if such a Hamiltonian were to be implemented as an actual computational device.

\emph{Tiling} problems, on the other hand, regard a set of square, edge-colored tiles, and ask whether it is possibe to tile the infinite plane with matching colors, where we are not allowed to rotate the tiles. This problem is known to be undecidable \cite{Robinson1971,gottesman2009quantum}, and is the foundation of various undecidability or uncomputability results on spectral properties of local Hamiltonians. Computation in these constructions is done similar to cellular automata: the ground state of such a Hamiltonian has a product structure, and is \emph{not} in a superposition over computational basis states.
In \cite{Bausch2015}, for instance, the authors embed a Turing Machine into a local Hamiltonian on a 2D lattice, and use one spatial direction of the system for its evolution in time, and the other to store the tape.
In that sense the state of cell at every point in time can be accessed by measuring a single spin.

One fundamental difference between Tiling and History State constructions is that with a Tiling Hamiltonian, we get a constant spectral gap. The problem with such Tiling Hamiltonians, however, is that it is not clear how to perform \emph{quantum} computation with it, and a no-go theorem for embedding unitary gates into a Tiling Hamiltonian \cite{AQIT2014} as well as the implications of such a gap amplification (we could solve the \emph{Quantum PCP Conjecture} if we could amplify to an increasing spectral gap, see e.g.~\cite{Vidick2016}) sound discouraging---but also challenging, as there is no known result forbidding this possibility.

We believe that there are a lot of unexplored problems in this field of research.
In particular, both the set of local interactions for History State constructions---sums of Hadamard-like operations---and Tiling interactions---diagonal projectors---only span a low-dimensional subspace over all possible interactions, and while interactions in general have been classified, it is not clear whether there exists a more optimal (in terms of the resulting promise gap, computational overhead, or precision required) one than History States, or not. A lot of research has been done to improve unitary gate usage in various contexts \cite{Aho2003,Abdessaied2014},
but we know little to nothing about reducing the overhead of embedding computation into the ground space of a Hermitian operator. For the hardness results, the constructed embeddings follow the usual \emph{Solovay-Kitaev} Karp reduction from a language \LL---complete for e.g.~\NP or \BQEXP---which results in a polynomial computation overhead.
The constructions themselves usually perform computation in a very localized fashion, meaning that there is some notion of heads, similar to a Turing Machine, while most of the space of the underlying Hilbert space remains a passive computation tape. This is not efficient: if we were able to more directly embed quantum circuits into a local Hamiltonian, we could drastically reduce the number of qubits needed to perform the calculation.

But using Hamiltonians to perform computation remains attractive:
adiabatic quantum annealing \cite{McGeoch2014} is the only area of experimental quantum computing claiming significant qubit counts\footnote{At the time of writing this article, small universal quantum devices with $\approx 50$ qubits are emerging, but they still lie above the error threshold required for extended computation \cite{Calude2017,Otterbach2017,Hsu2018}.}. By tuning the couplings between qubits, one hopes to encode a computation into the ground state of the Hamiltonian. Cooling and measuring the system then reveals the computation output.
The interactions, however, are typically limited to Ising-type couplings, and the physical interaction graph is fixed: logical interactions have to be crafted onto this underlying structure.
Furthermore, the coupling strengths have to be hand-tuned to represent a problem of interest, or calculated on classical hardware.

Despite these limitations, specialized hardware such as the \emph{D-Wave}---with their underlying \emph{Chimera} interaction graph (see e.g.\ \cite{Katzgraber2014})---has one significant advantage over a full-fledged analogue programmable quantum computer: we can build them today.

\emph{In this work, we want to focus not on embedding a general model of computation---Turing machines or circuits---into the ground space of Hamiltonians, but a very specific one: classifiers, similar to the ones we see in the context of neural networks.}

\subsection{From 3-SAT Tilings to Ground State Classifiers}
As a concrete and educational example for the type of question we address in this work, consider a quantum system with three spin-$1/2$ particles.
\[
\begin{tikzpicture}[
every node/.style={
	circle, draw=black, fill=black!20, minimum size=.5cm, inner sep=0pt
}
]
\draw (0, 0) node {$a$} -- (1, 1.7) node {$b$} -- (-1, 1.7) node {$c$} -- (0, 0);
\end{tikzpicture}
\]
We couple the particles with a three-local projector in the computational basis.\footnote{Note that, in this case, the interaction is in fact a global coupling.}
More concretely, we have available the interaction set
\[
S = \{
\ketbra{000}, \ketbra{001}, \ketbra{010}, \ldots, \ketbra{111}
\},
\]
from which we can build a Hamiltonian $\op H=\sum_{\op h\in S}a_{\op h} \op h$, for real coefficients $a_{\op h}$.
What we want to achieve is that the ground space of $\op H$ is spanned by vectors encoding the truth table of an AND gate, which can be achieved e.g.\ by labeling the first two qubits as ``input'', and the last qubit as ``output'', and choosing
\[
\op H = \ketbra{001} + \ketbra{010} + \ketbra{101} + \ketbra{110}.
\]
Invalid operations of the AND gate---such as $AND(0, 0)$ returning $1$---are penalized by the corresponding projector $\ketbra{001}$.
With this method, it is straightforward to encode a 3-SAT instance into the ground state of a Hamiltonian, as aforementioned.

While it was trivial to guess the coefficents for $\op H$ to encode AND, since we had 3-local interactions available, the story is different if we only allow \emph{2-local} interactions;
more specifically, we alter the interaction structure and only allow pairs of qubits $e$ to interact via the couplings
\[
S_e = \{ \ketbra{00}, \ketbra{01}, \ketbra{10}, \ketbra{11} \}.
\]
The overall Hamiltonian in this case is $\op H'=\sum_e\sum_{\op h\in S_e}a_{\op h,e}\op h^{(e)}$.
What are the optimal weights $a_{\op h,e}$ that give a ground state spanned by the truth table of AND? Is this possible, at all?

We might gain an advantage by going to a more elaborate two-local interaction graph, allowing for ``hidden'' spins, e.g.
\[
\begin{tikzpicture}[
every node/.style={
	circle, draw=black, fill=black!20, minimum size=.5cm, inner sep=0pt
}
]
\draw (0, 0) -- (4, 0) (1, 1.7) -- (3, 1.7);
\draw (0, 0) node {$a$} -- (1, 1.7) node[fill=white] {} -- (2,0) node {$c$} -- (3,1.7) node[fill=white] {} -- (4, 0) node {$b$};
\end{tikzpicture}
\quad\raisebox{1cm}{\!\!\!\text{or}\ }\quad
\begin{tikzpicture}[
every node/.style={
	circle, draw=black, fill=black!20, minimum size=.5cm, inner sep=0pt
}
]
\draw (0, 0) -- (4, 0) -- (4, 1.7) -- (0, 1.7) -- (0, 0) (2, 0) -- (2, 1.7);
\draw (0, 0) node {$a$} (2, 1.7) node {$b$} (4, 0) node {$c$};
\draw (4, 1.7) node[fill=white] {} (2, 0) node[fill=white] {} (0, 1.7) node[fill=white] {};
\end{tikzpicture}
\]
The white spins act as mediators for the correlations we require, but we allow them to be in any state in the ground space of $\op H'$, i.e.\ by only requiring that the marginal of the ground state obtained when tracing out all those ``hidden'' vertices is close to the truth table we require.
Another obvious modification is to choose a different set of interactions, e.g.\ a complete Pauli basis.
However it is not clear `a priori which graph works well with which set of interactions, and it is not clear how large the graph has to be to accurately model the target data (the Boolean truth table of AND in this example).

For 3-local interactions and e.g.\ a grid of arbitrary size we can model any classical logic circuit in the ground space, by writing a single 3-local diagonal projector per logic gate, as in our first example.
In that sense the model is complete for the complexity class \PP.

This question is naturally related to classifying restricted sets of interactions as in \cite{Piddock2015}, e.g.\ for commuting terms \cite{Bouland2016};
less obvious, for instance, is the fact that the local Hamiltonian problem remains \NP-hard, even if we allow commuting local terms and e.g.\ restrict to non-frustrated Hamiltonians with a constant promise gap \cite{Schuch2011,Aharonov2011}, or locally expanding graphs \cite{Aharonov2013}.

Yet even though this tells us something about the computational power of local Hamiltonians, there are a series of key differences to what we aim to achieve.
First of all, we are not necessarily interested in obtaining a universal computational model, but we want to obtain a \emph{classifier}, labelling data as \yes---for instance the valid AND applications---and \no---corresponding to the complement of the truth table.
The notion of hardness of estimating the ground state energy is fundamentally different, in the sense that we will have a single Hamiltonian encoding both \yes and \no instances which we can probe;
for the local Hamiltonian, the system with the couplings \emph{itself} is already the \yes or \no instance.

Furthermore, proof methods for a model's completeness, such as perturbation gadgets, often require two different energy scales for the local interaction strengths which furthermore grow with the system's size.

And finally, completeness for \PP does not suffice to give a good computational model,
not least due to the fact that the encoded circuits have to be produced by hand;
if we already knew a programmatical way of differentiating \yes and \no-labelled data beforehand, it would make little sense to write a Hamiltonian that performs this computation.

Indeed, our goal is to find a method which starts at the labelled data, graph, and an interaction set for every edge, and independently trains the coupling strengths such that the data can be discriminated as best as possible.

\subsection{Quantum Neural Networks and Ground State Classifiers}
Classification of data is a well-studied and empirically tested method which is used ubiquitously in our everyday lives.
Prominent examples where neural networks excel at classification tasks are in image recognition \cite{Karpathy2014}, or analysing game states to outperform human players at games like Go \cite{Silver2016}.
In a sense our model is a type of neural network, where we link nodes---e.g.\ the grey-shaded in- and output spins in our interaction graphs---with local coupling terms, which could connect the grey nodes over several layers of ``hidden'' spins.

The idea of extending neural networks to the quantum realm is not new, of course. An excellent overview over past efforts can be found in \cite{Schuld2014}. This task is not straightforward: measurements have to be included to introduce nonlinearity, which interferes with the coherence of the model, or post-selection is used to collapse a superposition of networks. A system of interacting quantum dots is suggested, capable of simulating network dynamics. However, experimentally, it is unclear how this is easier than building an analogue quantum computer.

We introduce a novel notion of ``quantum classifier'', based on the idea of finding a ground state of a local spin system with trained coupling strengths---e.g.\ quantum annealing---and probing it against some test data.
We want to point out that we regard a static property of a many-body quantum system (its low energy spectrum) and there are no dynamics (i.e.\ no forward propagation of data), in contrast to what is generally understood when quantum neural networks are mentioned.

Yet, to motivate this connection further, let us have a brief look at how a classical neural network could be set up to perform a classification task; for an extensive overview we refer the reader to \cite{Bishop2006}.

\begin{figure}
\begin{center}
\begin{tikzpicture}[
    every node/.style={
        circle, draw=black, minimum size=.5cm, inner sep=0pt
    }
]
\def\dx{2}
\def\dy{1.2}
\foreach \x/\X in {1/2,2/3,3/4}
    \foreach \a in {1,2,...,5}
        \foreach \b in {1,2,...,5}
            \draw (\dx*\x,\dy*\a) -- (\dx*\X,\dy*\b);

\foreach \a in {1,2,...,5}
    \draw[] (\dx*4,\dy*\a) -- (\dx*5,\dy*3);            
            
\foreach \y in {1,2,...,5}
    \draw (\dx*1,\dy*\y) node[fill=black!20] {$i_\y$};
\foreach \x in {2,3,4}
    \foreach \y in {1,2,...,5} {
        \draw (\dx*\x,\dy*\y) node[fill=white] {$f$};
    }
\draw (\dx*5,\dy*3) node[fill=black!20] {$o$};
\end{tikzpicture}
\end{center}
\caption{Feed forward classifier neural network with five input nodes, one output node, and three fully-connected hidden layers of size five.
Each line represents one (e.g.\ floating point) value being propagated forward from node to node; $f$ is a non-linear activator (e.g.\ sigmoid, ReLU) applied to an affine transformation on the input of the nodes; in general one allows $f$ to vary from layer to layer.}
\label{fig:classifier-nn}
\end{figure}
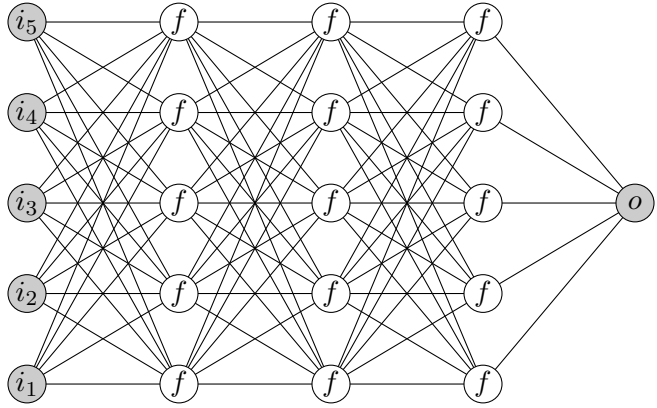
The input to our example is a string of numerical values representing the data.
For instance, an image could be translated into a series of pixels, each of which contains three floating point numbers with the RGB values.
Binary data (i.e.\ true or false values) can be mapped to the two values $0$ and $1$, and so on.
The network topology dictates how said values are processed: each node represents an activation function $f$ applied to an affine transformation of its input values: for any node $j$ in layer $k$, we have
\[
    \mathrm{out}^{(k)}_j = f\left(\sum_{i}W^{(k)}_{ij} * \mathrm{in}^{(k)}_i + b^{(k)}_j\right).
\]
Here, the $W^{(k)}$ and $b^{(k)}$ are weight matrices and bias vectors; the latter---and generally also $f$---can differ from layer to layer.

The output value at the right end of the network is then supposed to \emph{classify} the input data, by which we mean it should be as close as possible to the desired ``true'' label.
This can be achieved by training the network parameters $(W, b)$, e.g.\ using a variant of stochastic gradient decent (typically ADAM, see \cite{Kingma2014}) to optimize some loss function (e.g.\ cross entropy between obtained and expected output).

As an example, we could see the task of classifying integers into prime and not prime; the training data, if we represent the input digits in binary, would then be instances of the form
\begin{align}\label{eq:prime-data}
\Lyes &:= \{ 00101_2, 10001_2, \ldots \} \subset \mathds N_\mathrm{prime} \nonumber\\
\text{and } \Lno &:=\{ 10000_2, 00000_2, \ldots \} \subset \mathds N_\mathrm{prime}^c,
\end{align}
respectively.
The trained network's performance can then be benchmarked---ideally against previously unseen labelled data not in $\Lyes\cup\Lno$.

The type of classifier we hope to construct is of similar nature, but using the spectral properties of a local Hamiltonian $\op H$ in order to achieve the task, instead of a feed-forward neural net.
The goal is to construct $\op H$ such that its ground space has large overlap with the \yes instances, and little overlap with \no instances.

To explain this in more detail, consider a local Hamiltonian, which is a Hermitian operator of the form $\op H=\sum_i a_i \op h_i$, where the $\op h_i$ are at most $k$-local terms on a multipartite Hilbert space $\H=(\C^d)^{\otimes n}$---i.e.\ $\op h_i$ acts as identity on all but $k$ of the $d$-dimensional spins---and the $a_i$ are (real) coupling constants.
We denote the ground space of $\op H$ with $\gs(\op H)$.

The interaction structure of the many-body system can be represented with a multigraph $G=(V,E)$, which contains a multiedge $e_i$ for each term $\op h_i$: $e_i$ contains all those vertices that $\op h_i$ acts on non-trivially.
As an example, we can consider a spin chain with nearest neighbour interactions; $d$ is then the spin dimension (e.g.\ $d=2$ for qubits), the interactions are $k=2$-local, and $n$ is the spin chain length.
The corresponding interaction graph is a path graph of length $n$.
Another example of a valid interaction graph is given by the neural net in \cref{fig:classifier-nn}, which has two-local interactions between any ``spin'' in neighbouring layers.

\begin{figure}
\begin{center}
\begin{tikzpicture}[
    every node/.style={
        circle, draw=black, minimum size=.5cm, inner sep=0pt
    }
]
\def\dx{2}
\def\dy{1.2}
\foreach \a in {1,2,...,5}
    \draw (\dx*1,\dy*\a) -- (\dx*4,\dy*\a);
\foreach \b in {1,2,3,4}
    \draw (\dx*\b,\dy*1) -- (\dx*\b,\dy*5);

\foreach \a in {2,3,4}
    \draw[] (\dx*4,\dy*\a) -- (\dx*5,\dy*3);            
            
\foreach \y in {1,2,...,5}
    \draw (\dx*1,\dy*\y) node[fill=black!20] {$d_\y$};
\foreach \x in {2,3,4}
    \foreach \y in {1,2,...,5} {
        \draw (\dx*\x,\dy*\y) node[fill=white] {};
    }
\draw (\dx*5,\dy*3) node[fill=black!20] {$d_6$};
\end{tikzpicture}
\end{center}
\caption{Local spin system with a layout similar to \cref{fig:classifier-nn}, but with spins of interaction degree at most four.
Each line connecting two nodes represents a two-body interaction, which can differ from site to site. Five spins are arbitrarily marked as ``data'' spins, which can be seen as related to the in- and output nodes of a classical NN.}
\label{fig:classifier-ph}
\end{figure}
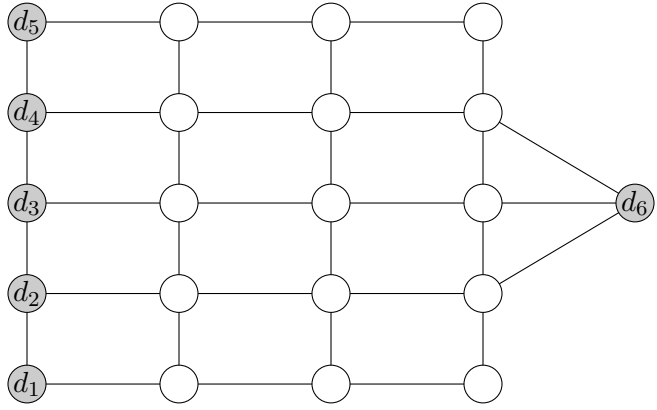

Observe, however, that---regarded as a many body system---this interaction topology is highly unrealistic: the graph cannot be embedded (without edges crossing) in, say, even three-dimensional space, and the degree of any vertex grows when the system size increases.
A more realistic variant with the same number and layout of vertices is given in \cref{fig:classifier-ph}, where the vertex degree is upper-bounded by a constant, in this case 4.

The idea is to take labelled data in binary and write them as kets; e.g.\ for our prime number example \cref{eq:prime-data}, we would have
\begin{equation}\label{eq:prime-data-quantum}
\Lyes' := \{ \ket{00101}, \ket{10001}, \ldots \} \text{ and } \Lno' :=\{ \ket{10000}, \ket{00000}, \ldots \}.
\end{equation}
Starting from a multigraph $G=(V,E)$ (e.g.\ the one in \cref{fig:classifier-ph}) and a finite set of $|E|$-local interactions $S_e$ for every edge $e\in E$ (note that we allow this set to vary from edge to edge),
we propose the following definition.
\begin{definition}[Classifier Hamiltonian, Variant A]\label{def:ham-A}
Take an interaction graph $G=(V,E)$ with a subset of marked vertices $V_d\subset V$, a family of edge sets $\{S_e\}_{e\in E}$, and labelled data $\Lyes,\Lno$.
Define $\Pi_\yes := \{ \ketbra{l}\otimes \1_{V_d^c} : l\in\Lyes \}$ as a projector onto \yes data acting non-trivially only on $V_d$, and extended everywhere else as $1$.

We say that the corresponding \emph{classifier Hamiltonian} $\op H$ is given by
\[
    \op H:=\sum_{e\in E}\sum_{\op h\in S_e}a_{e,\op h} \op h
\]
with parameters $a_{e,\op h}$ such that
\begin{enumerate}
\item the angle between the ground space of $\op H$ and the space spanned by the yes-instances $\sphericalangle (\gs(\op H),\supp \Pi_\yes)$ is maximized, and
\item the angle $\sphericalangle (\gs(\op H),\supp \Pi_\no)$ is minimized.
\end{enumerate}
\end{definition}

The data projectors act as identity everywhere but on the data vertices $V_d$, and as an identity operation on the complement $V_d^c$.
Therefore, the angle between ground space and \yes or \no data can be obtained by partially tracing out all but the data vertices first, i.e.
\[
\sphericalangle(\gs(\op H), \supp \Pi_{\yes/\no}) \equiv \sphericalangle\Big( \tr_h\gs(\op H),\!\!\! \sum_{l\in\mathrm L_{\yes/\no}} \!\!\!\ketbra l \Big).
\]
In analogy to neural network hidden layers we thus call the vertices in $V_d^c$ ``hidden'', denoted $V_h$ in the following; a partial transpose over hidden spins we write as $\tr_h(\cdot)$.

An alternative formulation for a classifier Hamiltonian is given in the following definition.
\begin{definition}[Classifier Hamiltonian, Variant B]\label{def:ham-B}
Take the same setup as in \cref{def:ham-A}, but choose the parameters $a_{e,\op h}$ such that
\begin{enumerate}
\item the energy expectation value $\tr(\op H \Pi_\yes)$ is minimized, and
\item the energy expectation value $\tr(\op H \Pi_\no)$ is maximized.
\end{enumerate}
\end{definition}

How can we hope that either of these models---optimizing ground state overlap in \cref{def:ham-A}, or optimizing energy expectation in \cref{def:ham-B}---has the ability to learn our data?
The idea is that the correlations present in the training data can be transferred to the spectrum of $\op H$.
This is not far-fetched: examples where long-range correlations emerge from local interactions are for instance Motzkin walks \cite{Movassagh2014}, where the ground state is highly degenerate and spanned by states corresponding to balanced open and close brackets, or simply a ferromagnetic Heisenberg model, where the spins in the ground state all align.
Similarly, we can hope that by tuning the interactions appropriately, we can create correlations between non-hidden spins that resemble the training data.

It is worth pointing out that the models we describe do not map one to one between neural networks and many-body spin systems, as in the latter we do not really have a notion of input and output vertices; we cannot label the \yes and \no data with an extra $\ket 1$ and $\ket 0$ acting on the spin $d_6$ in \cref{fig:classifier-ph}, for instance, and simply putting a large penalty on that spin to be in state $\ket 0$ will not work, as the previously unseen data does not have such a label.
It is therefore not clear `a priori whether an interaction structure similar to classical neural networks will yield a ``good'' classification performance.

\section{Main Results}
In this work, we present three main results.
\begin{enumerate}
\item We analyse whether \cref{def:ham-A,def:ham-B} are well-defined, and address the question whether they are suitable to perform non-trivial classification tasks.
\item We propose a general training scheme for the local coupling constants based on (simulated) quantum annealing, and test it  on small toy problems.
\item We benchmark interactions and graphs, and discuss possible implications of our empirical findings.
\end{enumerate}

The proposed training algorithm can be completely incorporated into an annealing device, avoiding the need for any classical optimization or repeated measurement to introduce non-linearity, and reducing unnecessary decoherence.
We construct a few such neural networks with \Learn---an F\# library based on \Liquid which can be used to build, train and test quantum classifiers, see \cite{Bausch2017a}---demonstrating empirically that the training algorithm works as intended.

Finally, we address a few related questions, such as which annealing parameters are optimal, and put it to the ``real-world'' classification task of distinguishing red and blue colors.

\subsection{Training Hamiltonian Classifiers}
We want to train a Hamiltonian classifier, which we model as a connected hypergraph $G=(V, E)$, where the hyperedges $E$ connect one or more vertices, each of which labels a qubit. The size of the edge $e\in E$ corresponds to the maximum locality of the interaction in the edge set $S_e$; our goal is to find the coupling strengths $a_{e,\op h}$ for all $e\in E$ and all $\op h\in S_e$, such that the conditions $(1)$ and $(2)$ in \cref{def:ham-A,def:ham-B} are satisfied, respectively.

\subsection{Optimization of Ground Space Overlap}\label{sec:opt-groundstate}
In \cref{def:ham-A}, the target is to
\begin{enumerate}
	\item maximize $\sphericalangle (\gs(\op H),\supp \Pi_\yes)$, and
	\item minimize $\sphericalangle (\gs(\op H),\supp \Pi_\no)$.
\end{enumerate}
It is clear that there is no hope of calculating $\gs(\op H)$ efficiently in the general case: as we have discussed in the introduction, even approximating the ground space energy is already a \QMA hard problem, and apart from some special cases we do not know how to impose that $\op H$ can have a constant spectral gap for all parameters, such that an algorithm like\ \cite{Aharonov2013} can be applied.

What we  propose instead is a general training algorithm based on (simulated) quantum annealing, which seems to work well empirically, at least on small samples (see \cref{sec:empirical}); we make no claims about its asymptotic runtime, or whether it converges to a configuration that actually optimizes \cref{def:ham-A}.

The idea is to use a so-called \emph{controlled interaction}, which---in analogy to a controlled unitary gate in a circuit---acts on an auxiliary spin that controls how strong said interaction has to act.
For every local interaction $\op h$, we introduce an extra ancilliary qubit $c$, and augment
\begin{equation}\label{eq:controlled-h}
\op h \mapsto \ketbra 1_c\otimes\op h = \begin{pmatrix}
0 & 0 \\ 0 & \op h
\end{pmatrix}.
\end{equation}
If $\op H=\sum_{i=1}^N a_i\op h_i$ is the Hamiltonian we want to train the weights $a_i$ for, we construct a training Hamiltonian with $N$ control spins, and define
\begin{equation}\label{eq:controlled-H}
\op H_c := \sum_i \ketbra 1_i \otimes \op h_i.
\end{equation}
To obtain a training algorithm from this controlled Hamiltonian, we propose a series of techniques.

\paragraph{One-Shot Optimization.}
This is a two-step process.
In the first step, we add a projector onto the \yes instances, and in the second step, we subtract a projector onto the \no instances; more concretely, we write
\begin{equation}\label{eq:method-1-ham}
\op H_1 := \op H_c - \delta\1\otimes \Pi
\end{equation}
where $\Pi=\Pi_\yes$ or $\Pi_\no$, respectively, and $\delta\ge1$ is a scaling parameter to compensate for the energy scale of the local couplings; abusing notation, we will assume that during training all local terms have norm $\| \op h \|\ll 1$ and set $\delta=1$.
To give a simple example, consider the case where we have three local interactions $\op h_1$ and $\op h_2$ which we wish to train the weights for.
Again sorting the two control qubits such that they precede the original Hilbert space, we have
\begin{align*}
\op H_1
&=\ketbra 1_{c_1} \otimes \1_{c_2}  \otimes (\1 + \op h_1)  +  \1_{c_1} \otimes \ketbra 1_{c_2}  \otimes (\1 + \op h_2) + \1_{c_1} \otimes \1_{c_2} \otimes \Pi \\
&=\left[\begin{array}{*{4}{@{}C{2.1cm}@{}}}
- \Pi &               &               &  \\
& \op h_1 - \Pi &               &  \\
&               & \op h_2 - \Pi &  \\
&               &               & \op h_1 + \op h_2 - \Pi
\end{array}\right].
\end{align*}
We thus see that the diagonal blocks contain all possible combinations of the local interactions $\op h_i$, and a copy of the projector $\Pi$.

We then run an annealing algorithm on $\op H_1$ in order to find its ground state, and then individually measure the control qubits.
Assuming for the moment that $\op H_1$ is non-degenerate, we easily see from its block structure that the ground state must be contained completely within one of its control blocks.
This means in particular that each of the control qubits is either completely in state $\ket 0$, or completely in state $\ket 1$---the outcome obtained by measuring each control qubit individually thus allows us to uniquely determine the corresponding block.
Uniqueness is lost if the ground space is degenerate, of course, and we will discuss this case in due course.

How do we interpret the resulting measurement outcome?
If we know that the block containing the ground state contains the interactions $\sum_{\op h\in M}\op h$, then that sum of interactions pushed the support of $\Pi$ up the least.
Observe that we allow the interactions to have arbitrary eigenvalues, so it is not clear \`a priori which one of the blocks wins.

We can thus determine which subset of interaction has the smallest overlap with $\Pi_\yes$---call this set $M_\yes$, and similarly with $\Pi_\no$---with set $M_\no$.
We claim that the final Hamiltonian $\op H=\sum_{\op h}a_{\op h}\op h$, where
\begin{equation}\label{eq:weights}
a_{\op h}:=\begin{cases}
1  & \text{if $\op h\in M_\yes\setminus M_\no$} \\
-1 & \text{if $\op h\in M_\no\setminus M_\yes$} \\
0  & \text{otherwise},
\end{cases}
\end{equation}
has large ground space overlap with \yes-instances, and small overlap with \no-instances.

We consider a concrete example.
Assume we have an initial interaction graph \tikz{\draw (0,0) -- (1,0); \draw[fill=black!20,radius=1.2mm] (0, 0) circle[] (1,0) circle[]; }, i.e.\ two qubits, none of which are hidden.
The possible data set for this model is two bits, i.e.\ $\{\ket{00}, \ket{01}, \ket{10}, \ket{11}\}$.
Pick $\Lyes=\{ \ket{01},\ket{10} \}$, and $\Lno=\{ \ket{00}, \ket{11} \}$---which we could for instance interpret as a NOT-gate.

Assuming the list of interactions is $S=\{ \ketbra{00}, \ketbra{01}, \ketbra{10}, \ketbra{11} \}$,
it is straightforward to guess the correct Hamiltonian for this dataset: constraining the prefactors $a_i$ to the interval $[0,1]$, the optimal Hamiltonian in the sense of \cref{def:ham-B} is $\op H_\text{target}=\ketbra{00} + \ketbra{11}$.

Can we learn these parameters with our proposed method?
As in \cref{eq:method-1-ham}, and for the \yes step, we have the diagonal matrix
\newcommand{\Vd}{\tikz{\draw[dotted] (0,0) -- (0,.5);}}
\newcommand{\MM}{\underline{10}}
\[
\begin{array}{l *{20}{@{}C{.5cm}@{}}}
\op H_1 = -\Big[ & 0  & \MM & \MM & 0 & \Vd & 0  & \MM & \MM & -1 & \Vd & 0  & \MM & 9 & 0 & \Vd & 0  & \MM & 9 & -1 &   \Vd    \\
& 0  & 9  & \MM & 0 & \Vd & 0  & 9  & \MM & -1 & \Vd & 0  & 9  & 9 & 0 & \Vd & 0  & 9  & 9 & -1 &   \Vd    \\
& -1 & \MM & \MM & 0 & \Vd & -1 & \MM & \MM & -1 & \Vd & -1 & \MM & 9 & 0 & \Vd & -1 & \MM & 9 & -1 &   \Vd    \\
& -1 & 9  & \MM & 0 & \Vd & -1 & 9  & \MM & -1 & \Vd & -1 & 9  & 9 & 0 & \Vd & -1 & 9  & 9 & -1 & \ \Big].
\end{array}
\]
The ground space of $\op H_1$ is thus degenerate and spanned by vectors with support in all but four the blocks (containing $-10$, underlined).
The control measurement probabilities for the \yes and \no steps are shown in \cref{tab:measurement-1,tab:measurement-2}, respectively.

\begin{table}[th]
    \centering
	{\begin{tabular}{clll}
			\toprule
			control \# & interaction   & $\tr(\cdot \ketbra1)$ & shifted$^\mathparagraph$ \\ \midrule
			1      & $\ketbra{00}$ & $0.25$                & 0                  \\
			2      & $\ketbra{01}$ & $0.75$                & 1                  \\
			3      & $\ketbra{10}$ & $0.75$                & 1                  \\
			4      & $\ketbra{11}$ & $0.25$                & 0                  \\ \bottomrule
	\end{tabular}}
	\caption{Measurement probabilities for control qubits for toy example \yes training step. $^\mathparagraph$Probabilities normalized to the interval $[0,1]$.\label{tab:measurement-1}}
	
	\vspace{3mm}
	{\begin{tabular}{clll}
			\toprule
			control \# & interaction   & $\tr(\cdot \ketbra1)$ & shifted$^\mathparagraph$ \\ \midrule
			1      & $\ketbra{00}$ & $0.75$                & 1                  \\
			2      & $\ketbra{01}$ & $0.25$                & 0                  \\
			3      & $\ketbra{10}$ & $0.25$                & 0                  \\
			4      & $\ketbra{11}$ & $0.75$                & 1                  \\ \bottomrule
	\end{tabular}}
	\caption{Probabilities for \no training step. $^\mathparagraph$Probabilities normalized to the interval $[0,1]$.\label{tab:measurement-2}}
\end{table}
In order to interpret the control measurements, it suffices to subtract the \yes from the \no weights, and then re-shift them to lie within the desired interval.
The resulting Hamiltonian is then precisely $\op H_\text{target}$, as intended.

If we consider the same example, but with an interaction comprising three random diagonal entries, there will in general be precisely \emph{one} ground state, and the resulting Hamiltonian will then necessarily be a linear combination of the interactions, but with coefficients either $-1$, $1$ or $0$, as in \cref{eq:weights}.

That this cannot generally be the optimal solution should be apparent, and we discuss the possible caveats in \cref{sec:empirical}.
One immediate consequence is that this training method does not pick out the interactions which have as little overlap with $\Pi$ as possible, but those where at least a \emph{single} data point yields the most negative eigenvalue.
In our empirical analysis, we did not find this to be an issue, but we want to propose an alternative training method which does not have the same problem.

\paragraph{Serial Optimization.}
The second proposed method is similar to the previous technique, but with the modification that we run one optimization step for each data point $l\in\LL$.
The data projectors $\Pi_\no$ and $\Pi_\yes$ are now replaced with rank one projectors $\ketbra{l}$; for each $l$, we collect which blocks are optimal as the family of sets $M_l$, as before.

The final weights can then be determined by a weighted count over the \yes and \no sets:
for all local terms $\op h$, we set the coefficients similar to \cref{eq:weights} as
\begin{equation}\label{eq:weights-serial}
a_{\op h} :=\frac{|\,
	\{
	\op h \in M_{l} : l\in \Lyes
	\}\,|}{|\Lyes|}
-\frac{|\,
	\{
	\op h \in M_{l} : l\in \Lno
	\}
	\,|}{|\Lno|}.
\end{equation}
This procedure has the advantage of reducing the locality of the data projector to 1-local (!) instead of $|l|$-local (which might be infeasible), but at the cost of having to run $|\LL|$ separate optimization steps.
Furthermore, we obtain a more accurate approximation to which combination of local interactions is optimal;
as aforementioned, if we train in one shot, the block with the smallest energy state is the one where a single combination of data vector and interactions gives the smallest penalty, effectively ignoring how the interactions fare on all other data points.
With serial optimization, we know how the interactions behave on \emph{all} data points.

\subsubsection{Optimization of Energy Expectation}\label{sec:opt-energy}
\paragraph{Exact Optimization.}
In variant B, which we expect to be the significantly simpler case, our aim is to
\begin{enumerate}
\item minimize $\tr(\op H\Pi_\yes)$, and
\item maximize $\tr(\op H\Pi_\no)$.
\end{enumerate}
Assuming without loss of generality that the hidden system $V_h$ is sorted to the back, we get
\begin{align}
    \tr(\op H\Pi_\yes)
    &= \tr\left(\sum_{e\in E}\sum_{\op h\in S_e}a_{e,\op h} \op h * \sum_{l\in\Lyes}\ketbra l\otimes\1_h \right) \nonumber \\
    &= \sum_{e\in E}\sum_{\op h\in S_e}\sum_{l\in\Lyes}a_{e,\op h} \tr( \op h * \ketbra l\otimes\1_h) \nonumber \\
    &= \sum_{e\in E}\sum_{\op h\in S_e}\sum_{l\in\Lyes}a_{e,\op h} \underbrace{\tr( \tr_h(\op h) * \ketbra l )}_{=:c_{e,\op h}}. \label{eq:tr-interactions}
\end{align}
A similar calculation can be done for $\Lno$;
given we want to restrict the parameters $a_{e,\op h}$ to lie within a certain range, e.g.\ in $[-1,1]$, we can express the problem of minimizing and maximizing conditions $(1)$ and $(2)$ as a linear program, which is efficient to solve as long as we can calculate the constants $c_{e,\op h}$ efficiently;
and indeed, since the locality of the $\op h$ is fixed in this case, \cref{eq:tr-interactions} can be calculated in poly-time in the size of the instance $l$, the number of vertices, and the number of interactions per edge, as can be verified easily.

\paragraph{Empirical Optimization.}
The second method of training we propose is a slight modification of $\op H_1$ in \cref{eq:method-1-ham}.
The reason for this proposition is to get an intuition on its performance on an annealing device, and not to improve on the exact optimization technique, which we already know to be efficient.
What we want to achieve is that the weight is concentrated in the diagonal block which, on average, gives the largest bonus to the data projector $\Pi$.
Since $\Pi$ is not a local coupling in general, we do not loose much by the following modification, namely
\begin{equation}\label{eq:method-2-ham}
\op H_2 := (\1 \otimes \Pi) \op H_c (\1 \otimes \Pi).
\end{equation}
In an example with two possible couplings to be trained, $\op h_1$ and $\op h_2$, we can calculate
\[
\op H_1
= \left[\begin{array}{*{4}{@{}C{2.1cm}@{}}}
0 &               &               &  \\
& \Pi\op h_1\Pi &               &  \\
&               & \Pi\op h_2\Pi &  \\
&               &               & \Pi(\op h_1 + \op h_2)\Pi
\end{array}\right].
\]
Again assuming non-degeneracy, we see that the smallest eigenvalue will be in whatever combination of couplings gives the highest bonus to the terms in the data projector; but if we want to have an average, we can trace out all but the control qubits.
This yields
\[
\tr_{\text{system}}\op H_1 = \diag\Big[
0 ,  \tr(\Pi\op h_1) ,  \tr(\Pi\op h_1) , \tr(\Pi(\op h_1 + \op h_2))
\Big].
\]
The ground state will thus be in the block where a combination of interactions has the largest energy bonus within the subspace spanned by the data.

Again we note that this method will not, generally, give the optimal solution; the problem is that although we know which combination of interactions is optimal, we do not know their weight.
The correct procedure in this case is thus to iterate the algorithm by learning which block is optimal, and then to increase the weights of the Hamiltonian terms included in the block by a small amount, in order to obtain an expression closer to the exact weights.

A similar two-spin example as we did for Method 1 shows, however, that especially for projectors this does not seem to be an issue, as the resulting ground space in the control qubits is highly degenerate, and therefore already captures all optimal subsets of interactions in the first step.
However, there is one caveat: tracing out the system $\tr_{\text{system}}$ could be a very inefficient procedure, depending on how it is implemented.

\section{Empirical Results}\label{sec:empirical}
\subsection{Implementation Details}
While it is feasible to diagonalize small Hamiltonians exactly, the ultimate goal is to perform the Hamiltonian classifier training on an annealing device itself.
Microsoft's software package \Liquid \cite{Wecker2014} allows us to simulate quantum annealing on classical hardware, with the obvious restrictions on memory requirements;
the annealer further has a hard-coded limit of $23$ qubits.
We developed a software library to interface \Liquid---\Learn  \cite{Bausch2017a}---which enables us to train and test Hamiltonian classifiers in a straightforward fashion.
Due to performance considerations, the training algorithm implemented and employed in the following is the one-shot optimization of ground state overlap from \cref{sec:opt-groundstate}.

In order to train a Hamiltonian classifier as given in \cref{eq:method-1-ham}, \Liquid allocates a \texttt{Ket} with one qubit per unique graph vertex, using \texttt{Liquid.Spin}, and we run the \texttt{Spin.Test} annihilation procedure, using the qubit interactions defined above as custom \texttt{SpinTerm}s. Internally, we run the simulator twice: once for the \yes instances and once for \no, as outlined in \cref{sec:opt-groundstate}. In every run, we assign an energy bonus to the data qubits in any state specified by the training dataset, and measure the control qubits using \texttt{Spin.EnergyExpectation}.

The final test assessment is then made by assembling the Hamiltonian $\op H$ using the trained weights;
we perform a complete (all basis states) joint measurement of the ground state on the output qubits to compare with the test data, and also calculate the energy expectation of the test state with respect to $\op H$, again using \texttt{Spin.EnergyExpectation}---the reason for the latter is that while ground state overlap is what we want to optimize, we only get a binary answer, i.e.\ how much the test state lies within the ground space.
To gain further insight into the energy spectrum of the trained Hamiltonian $\op H$, it will be useful to see how far up the energy spectrum the test state lies.
For implementation details, we refer the reader to the source code repository \cite{Bausch2017a}.

\texttt{Spin.Test} utilizes simulated quantum annealing to move from a known ground state of a trivial initial Hamiltonian with an efficiently-preparable ground state
\[
\op H_\text{init} = -\sum_{i=1}^n \sigma_x^{(i)}
\quad\text{with}\quad
\ket{\psi_\text{init}} = \frac{1}{\sqrt{2^n}}\sum_{s\in\{0,1\}^n}\ket s
\]
to the ground space of our target Hamiltonian $\op H_{1/2}$, by interpolating between them via
\[
\op H(t) = (1-t)\op H_\text{init} + t\op H_{1/2}
\quad\text{for}\ 
0\le t\le 1.
\]
If $t$ is varied sufficiently slowly---which depends on the minimal spectral gap of $\op H(t)$ along its path---the system remains close to its ground state for all $t$ (see e.g.\ \cite{McGeoch2014});
the final ground state of the target Hamiltonian can thus be accessed at the end of the annealing procedure.

\begin{figure}[t]
\begin{tikzpicture}[
    x=\columnwidth/120,
    y=\columnwidth/120/1.618,
    pct/.style={black!20}
]
\foreach \p in {0,25,50,75,100}
{
    \draw[pct] (-3, \p) -- (103, \p);
    \node[left] at (0, \p) {$\p\%$};
}

\draw[pct] (0, -3) -- (0, 103);
\draw[pct] (16.67, -3) -- (16.67, 103);
\draw[pct] (50, -3) -- (50, 103);
\draw[pct] (100, -3) -- (100, 103);
\node[below] at (0, 0) {$0$};
\node[below] at (16.67, 0) {$1/8$};
\node[below] at (50, 0) {$1/2$};
\node[below] at (100, 0) {$1$};

\draw[->] (0,0) -- (105, 0);
\draw[->] (0,0) -- (0, 105);

\draw[dashed] (0, 100) -- (16.67, 100) -- (100, 0);
\draw[red, dashdotted] (0, 0) -- (16.67, 25) -- (100, 100);
\draw[blue] (0, 0) -- (16.67, 0) -- (100, 100);
\node at (35, 84) {$\sigma_x$};
\node[red] at (25, 38) {$\Pi$};
\node[blue] at (75,61) {$\op H_c$};

\node[right, rotate=-270] at (3,76) {\contour{white}{strength}};
\node[above] at (100,0) {\contour{white}{time $t$}};

\end{tikzpicture}
\caption{Annealing schedule for the \Liquid simulation. The dashed line is the $\sigma_x$ interaction, the dash-dotted red line is the data penalty terms $\1\otimes\Pi$ and the solid blue line is the control Hamiltonian $\op H_c$ which we want to train.}
\label{fig:annealing-schedule}
\end{figure}
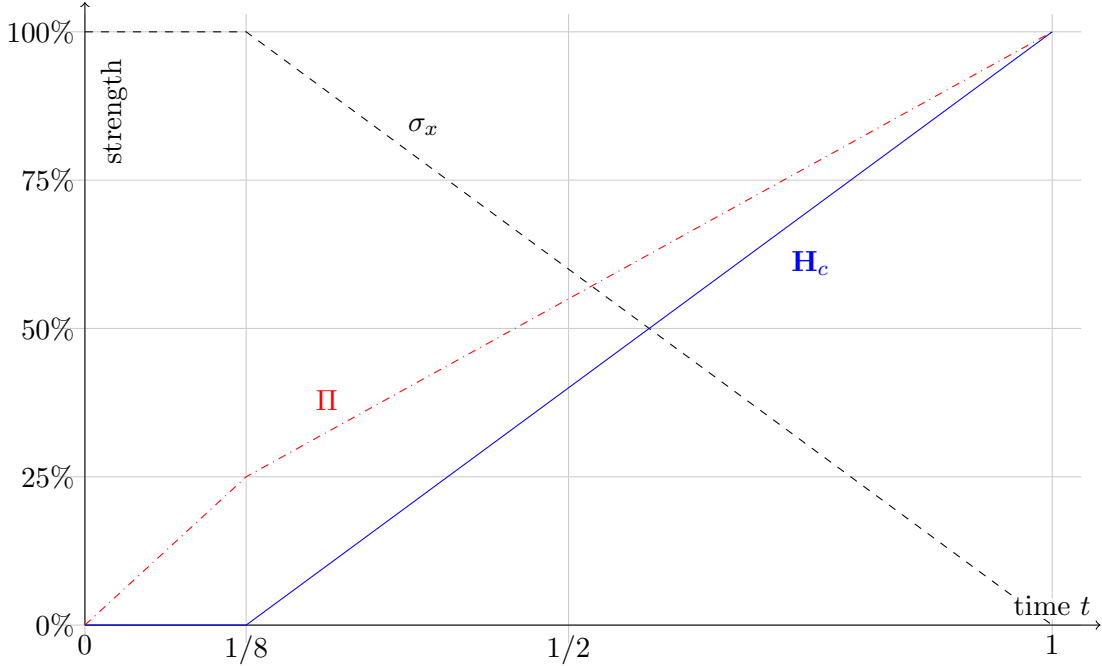

For our purposes, we found it useful to split the annealing procedure for \cref{eq:method-1-ham} into three paths, which are depicted in \cref{fig:annealing-schedule}:
\begin{enumerate}
\item $\op H_\text{init}$ kept at full strength within $t=[0,1/8]$, and then ramped down linearly for $t\in[0,1]$.
\item $\1\otimes\Pi$ is ramped up to $1/4$ of its strength for $t\in[0,1/8]$, and then increased linearly to full strength for $t\in[1/8,1]$.
\item $\op H_c$ is kept at zero strength for $t\in[0,1/8]$, and then ramped up linearly for $t\in[1/8,1]$.
\end{enumerate}
The intuition behind this annealing schedule is that within the first section of the procedure, the ground space, initially in a uniform superposition over all bit strings, is ``filtered'' to be a uniform superposition over all data words.
The control Hamiltonian $\op H_c$ will therefore immediately see the potential landscape and the correlations induced by the training data, which restricts the effective dimension that $\op H_c$ has to roam.\footnote{We want to point out that this argument is based on intuition and empirical evidence, not on rigorous theoretical reasoning.}

\Liquid itself is a circuit simulator, and it therefore uses a method called \emph{trotterization} to simulate the annealing procedure (see \cite{Lloyd1996}).
The idea is that if we know that a system is governed by a local Hamiltonian
$
    \op H = \sum_{i=1}^k \op h_i
$,
one can approximate this evolution by a truncation of the Trotter product formula \cite{Trotter1959},
\begin{equation}\label{eq:trotter}
    \exp(\ii \op H t) = \left(\prod_{i=1}^k \exp(\ii \op h_i t/n_T)\right)^{\!\!n_T} + R_{n_T+1},
\end{equation}
where the remainder $R_n\rightarrow 0$ for $n\rightarrow\infty$.
This introduces a new meta-parameter $n_T$, and we benchmark in \cref{sec:e-benchmark-parameters} what influence the subdivision count has on the performance of the classifier.

A bit less obvious is the fact that we also have to choose a time resolution, as we can only run the trotterized quantum circuit for a discrete set of times $t\in[0,1]$;
we therefore subdivide the annealing interval $[0,1]$ into $R$ sections, i.e.\ $\{0,1/R,\ldots,1\}$.
For each time step $t$, we calculate a unitary quantum gate for all interactions $\op h_i$ by calculating its matrix exponential.

We want to give an example, which emphasizes why our model of controlled interactions appears work.
Consider a local coupling $\op h$ which we want to train.
Calculating the matrix exponential of the corresponding controlled interaction
\[
    \op U(\ketbra{1}_c \otimes \op h) = \exp\left(\ii t \begin{pmatrix}
    0 & 0 \\ 0 & \op h
    \end{pmatrix}\right) = \begin{pmatrix}
    \exp(\ii t 0) & 0 \\ 0 & \exp(\ii t \op h)
    \end{pmatrix} = \begin{pmatrix}
    \1 & 0 \\ 0 & \op u
    \end{pmatrix},
\]
which is precisely a controlled-$\op u$ gate; in circuit notation, where $\ket c$ denotes the control spin, and $\ket\psi$ the system $\op h$ acts on,
\begin{center}
\begin{tikzpicture}[ x=1cm, y=-1cm ]
\draw[fill=black] (0, 0) -- (3, 0) (0, 1) -- (3, 1) (2, 0) circle[radius=.0707] -- (2, 1);
\draw[fill=white] (2, 1) +(-.3, -.3) rectangle +(.3, .3) +(0, 0)node {$\op u$};
\draw (-.25,0) node[left] {$\ket c$} (-.25,1) node[left] {$\ket{\psi}$};
\draw (3.25,1) node[right] {$\braket{0}{c}\ket\psi + \braket{1}{c}\op u\ket\psi$};
\end{tikzpicture}
\end{center}
If $\ket c$ is close to $\ket 1$, $\op u$ is applied with full strength, corresponding to a stronger coupling constant in front of $\op h$ to leading order, since
\begin{align*}
\braket{0}{c}\ket\psi + \braket{1}{c}\op u\ket\psi 
& = \left[\braket{0}{c}\exp(\ii t 0) + \braket{1}{c}\exp(\ii t \op h) \right] \psi \\
& \approx \left[\braket{0}{c}(\1 + 0) + \braket{1}{c}(\1 + \ii t \op h) \right] \psi \\
& = \left[ \braket{0}{c} + \braket{1}{c} \right] \psi + \braket{1}{c} \ii t\op h \psi,
\end{align*}
where we assumed that $\op h = \1 + \mathrm{O}(1/n_T)$ is close to the identity, as prescribed by \cref{eq:trotter}.
We can thus see that in the language of local Hamiltonian couplings, a controlled interaction is raised to a controlled gate in a simulated annealing circuit,
and the control vertex strength translates to how strong said coupling is felt.

We note that due to the block structure of \cref{eq:controlled-H} there should never be any entanglement between the control qubits (as we measure them individually), beyond what we get due to degenerate blocks.
Measuring the control qubits individually should therefore be lossless, in the sense that the resulting trained Hamiltonian is expected to perform equally well as the training Hamiltonian did.

\subsection{Color Classification}\label{sec:e-color}
We would like to assess our setup with a relevant, ``real-world'' problem. Unfortunately, even the smallest examples from well-known machine learning datasets require way too many qubits as input to run on the quantum simulator, so we have to construct a somewhat artificial dataset which works within the hardware and software limitations at hand.\footnote{As aforementioned, \Liquid allows us to simulate 23 qubit quantum circuits. Since every graph edge with interaction set $S_e$ requires $|S_e|$ additional control vertices, this limit is reached quickly. We discuss in \cref{sec:optimization} how to reduce the number of required control vertices.}

We consider the task of labelling colors by the two classes \emph{red} and \emph{blue}. Our training dataset will consist of a list of 9 bit, little-endian color codes in RGB order, e.g.~001~100~111, which would represent a light blue tone (little red, half green, full blue).
This is an interesting, non-trivial task, as there are a lot of edge cases, as well as some colors which are neither considered red or blue---so let us see what our quantum classifier thinks.

\begin{figure}[t]
\centering
\begin{tikzpicture}[
    vertex/.style={
        circle, draw=black, minimum size=.5cm, inner sep=0pt
    }
]

\draw (-2, 2) node[] {a)};
\foreach \a/\t in {1/r_1,2/r_2,3/r_3,4/g_1,5/g_2,6/g_3,7/b_1,8/b_2,9/b_3}{
    \draw (\a*360/9: 2cm) -- (0, 0);
    \draw (\a*360/9: 2cm) node[vertex,fill=black!20] {$\t$};
}
\draw (0, 0) node[vertex,fill=white] {};

\begin{scope}[xshift=6cm]
    \draw (-2, 2) node[] {b)};
    \foreach \a/\t in {1/r_1,2/r_2,3/g_1,4/g_2,5/b_1,6/b_2}{
        \draw (\a*360/6: 2cm) -- (0, 0);
        \draw (\a*360/6: 2cm) node[vertex,fill=black!20] {$\t$};
    }
    \draw (0, 0) node[vertex,fill=white] {};
\end{scope}
\end{tikzpicture}
\caption{Interaction graph for (a) 9 bit, and (b) 6 bit color classification task, with two-local couplings and one hidden center node. Each edge carries as coupling a linear combination of the four possible two-local projectors $\ketbra{00},\ketbra{01},\ketbra{10},\ketbra{11}$.}
\label{fig:9color-graph}
\end{figure}
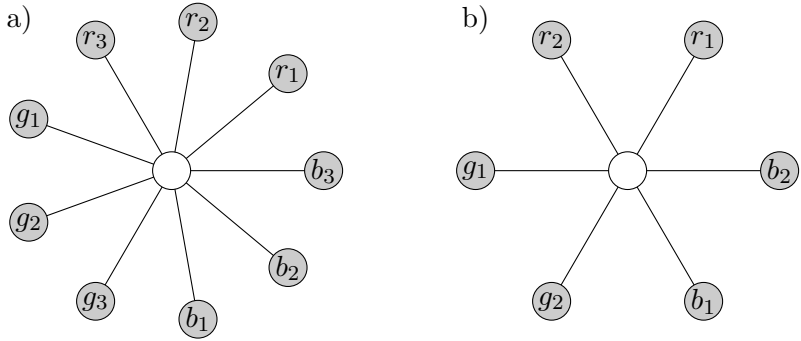

The graph we will train this example on can be seen in \cref{fig:9color-graph}.
We use 2-local projectors on computational basis states as interaction set on every edge; the overall Hamiltonian will thus be
\begin{equation}\label{eq:9color-H}
\op H = \sum_{c\in\{r, g, b\}}\sum_{i=1}^3(a^{00}_{c,i}\ketbra{00} + a^{01}_{c,i}\ketbra{01} + a^{10}_{c,i}\ketbra{10} + a^{11}_{c,i}\ketbra{11}),
\end{equation}
and our goal is to obtain the optimal coupling constants in order to optimize the two overlap conditions in \cref{def:ham-A}.

\begin{table}[b]
\centering
{\begin{tabular}{l|cl}
	\toprule
	instance                               &                       colors                        & binary representation$^\mathparagraph$                \\ \midrule
	9 bit, red                             &   \includegraphics[width=6cm]{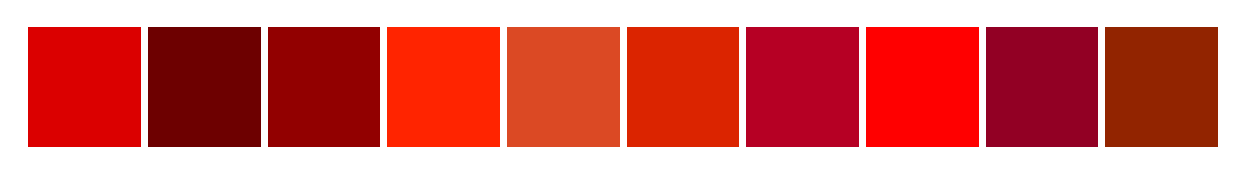}    & \makecell[l]{$\ket{110000000}$, $\ket{011000000}$, \\
	$\ket{100000000}$, $\ket{111001000}$,    \\
	$\ket{110010001}$, $\ket{110001000}$,    \\
	$\ket{101000001}$, $\ket{111000000}$,    \\
	$\ket{100000001}$, $\ket{100001000}$.}   \\ \midrule
	9 bit, blue                            &   \includegraphics[width=6cm]{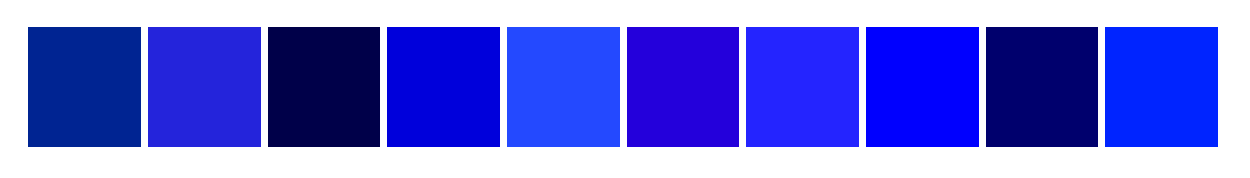}   & \makecell[l]{$\ket{000001100}$, $\ket{001001110}$, \\
	$\ket{000000010}$, $\ket{000000110}$,    \\
	$\ket{001010111}$, $\ket{001000110}$,    \\
	$\ket{001001111}$, $\ket{000000111}$,    \\
	$\ket{000000011}$, $\ket{000001111}$.}   \\ \midrule
	6 bit, red                             & \includegraphics[width=3.2cm]{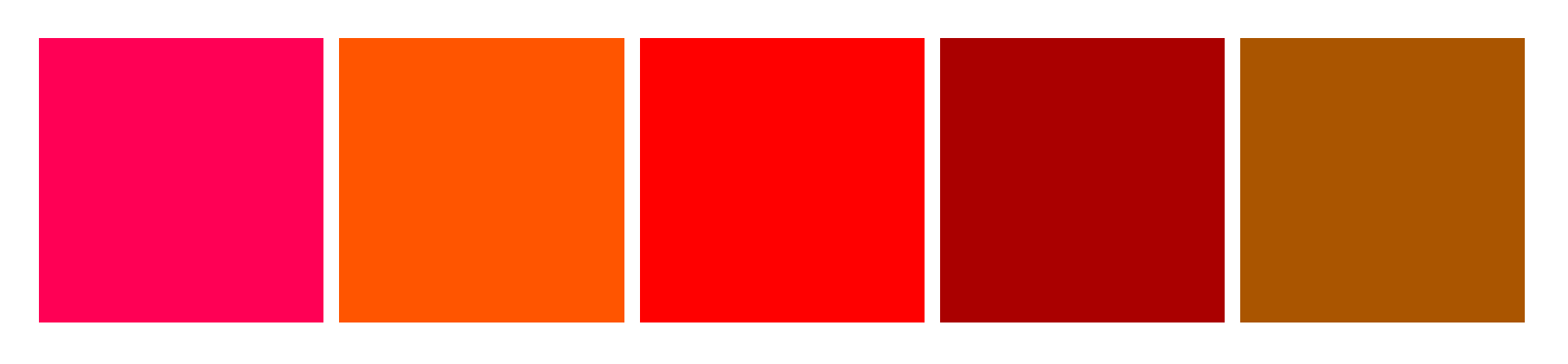}  & \makecell[l]{$\ket{110001}$, $\ket{110100}$,       \\
	$\ket{110000}$, $\ket{100000}$,          \\
	$\ket{100100}$.}                         \\ \midrule
	6 bit, blue                            & \includegraphics[width=3.2cm]{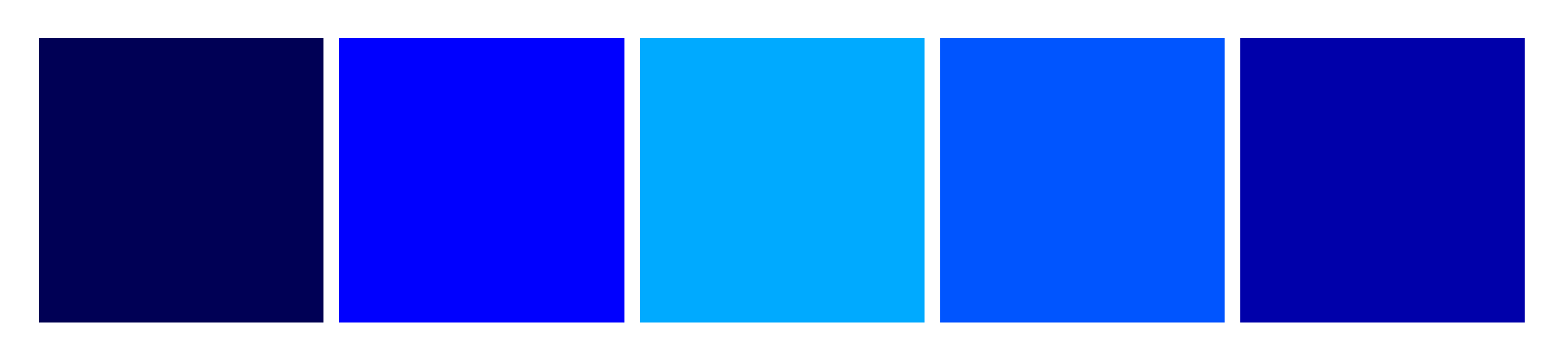} & \makecell[l]{$\ket{000001}$, $\ket{000011}$,       \\
	$\ket{001011}$, $\ket{000111}$,          \\
	$\ket{00010}$.}                          \\ \bottomrule
\end{tabular}}
\caption{Training data for 9 bit color classification task. $^\mathparagraph$In little endian RGB order.\label{tab:9color-training-data}}
\end{table}

\begin{figure}[bt]
\centering
\includegraphics[width=0.7\linewidth]{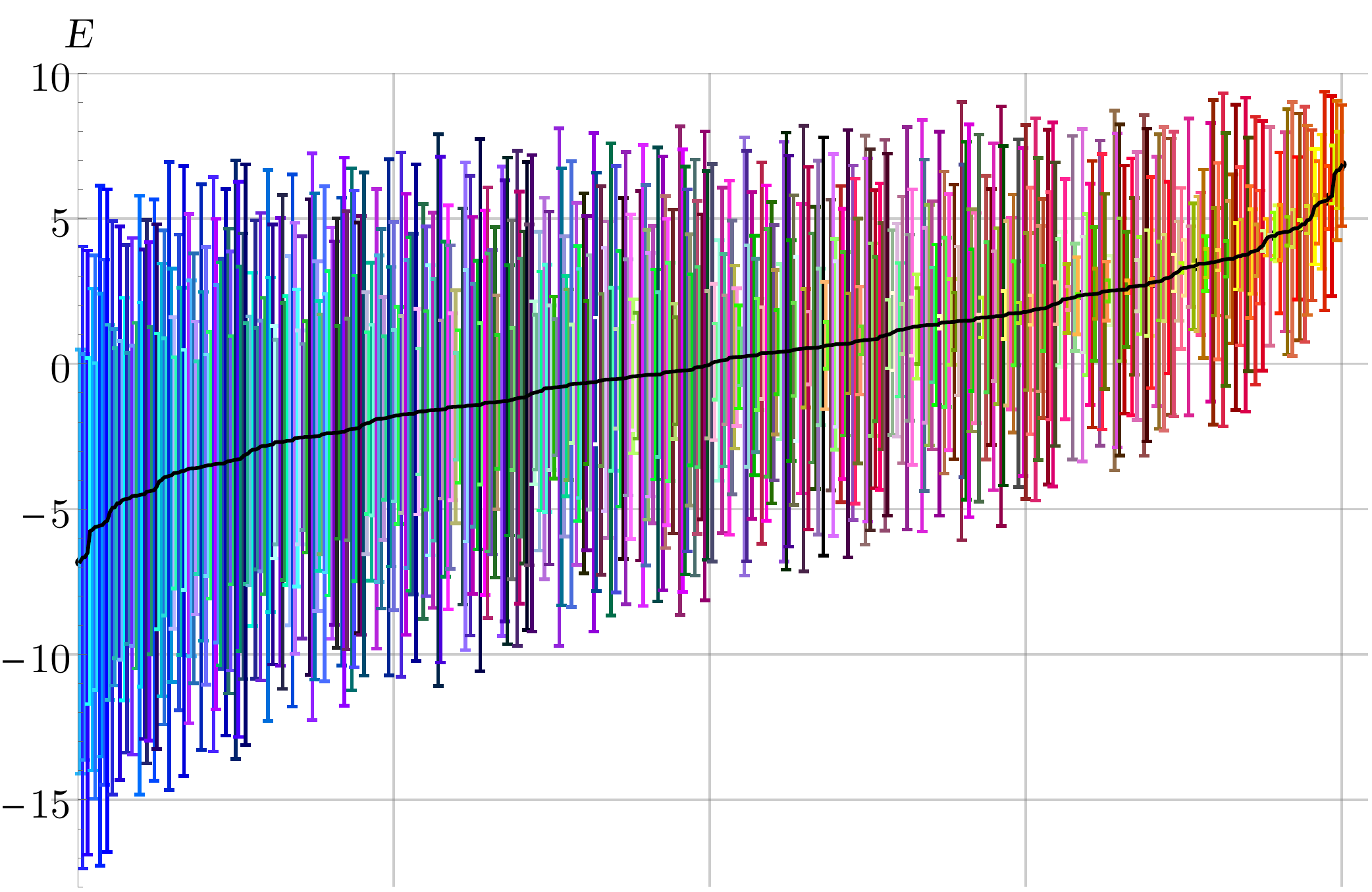}
\caption{Measured energy expectation and standard deviation as returned from \Liquid's \texttt{Spin.EnergyExpectation} for all 512 9-bit colors with respect to the ground space of the trained color classification Hamiltonian in \cref{eq:9color-H}, sorted by the expectation value.}
\label{fig:9color-result-1}
\end{figure}

\begin{figure}[bt]
\centering
\includegraphics[width=0.7\linewidth]{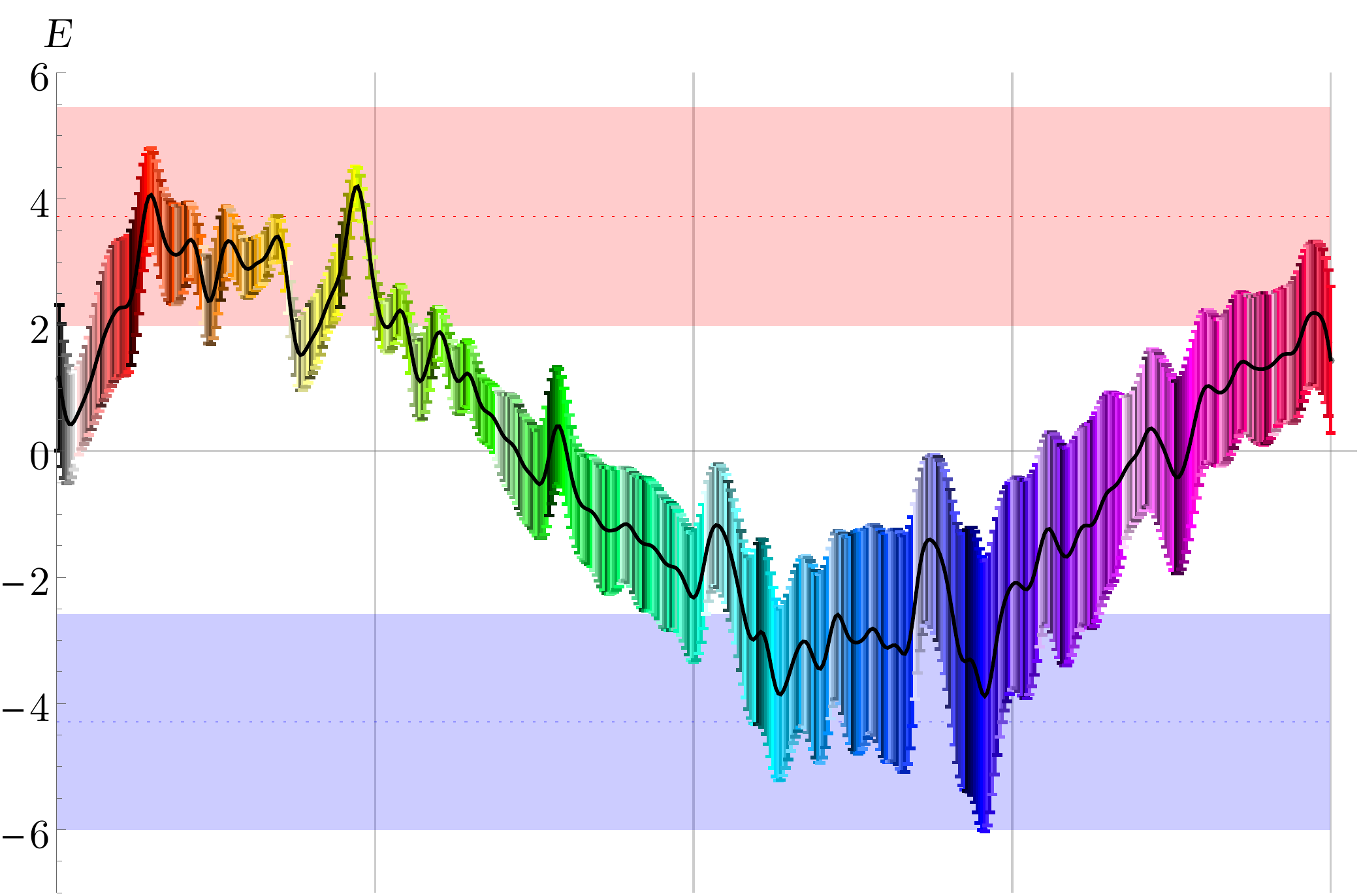}
\caption{Same data as in \cref{fig:9color-result-1}, but sorted by hue, and with a moving average (Gaussian kernel with a standard deviation of 3). The blue and red dotted lines mark the mean of the blue and red training data from \cref{tab:9color-training-data}, respectively, as well as their standard deviation (shaded areas).}
\label{fig:9color-result-2}
\end{figure}

Ten red and blue color shades are chosen randomly to train the network (shown in \cref{tab:9color-training-data}), and we test it on the full possible 512 color spectrum available with 9 bits. The output is plotted in \cref{fig:9color-result-1,fig:9color-result-2}. The error bars show the standard deviation of the energy expectation as returned by \Liquid's \texttt{Spin.EnergyExpectation}.

\begin{figure}[bt]
\centering
\hspace{-4mm}
\begin{tikzpicture}[]
\node (0, 0) {\includegraphics[width=0.71\linewidth]{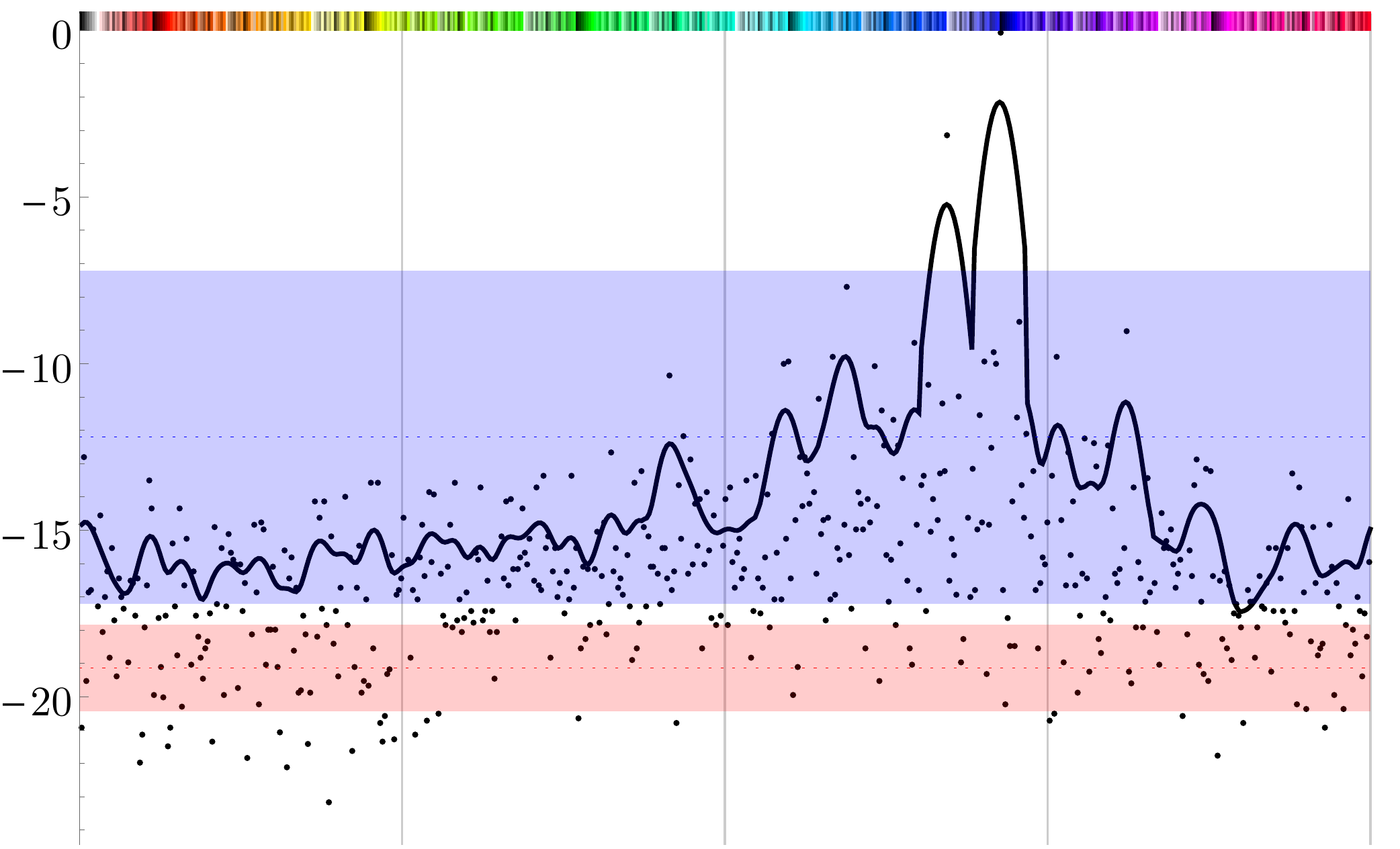}};
\draw (-4.28,-2.4) node {\small $\log p$};
\end{tikzpicture}
\caption{Distribution of probabilities $p$ of joint ground state measurement on all 9 output qubits, logarithmic scale. The solid line is a moving average with the same Gaussian kernel as in \cref{fig:9color-result-2}. The blue and red dotted lines mark the mean of the blue and red training data and their standard deviation (shaded areas).}
\label{fig:9color-result-3}
\end{figure}

The outcome shows a clear distinction between red and blue hues.
More specifically, blue and turquoise tones have the lowest energy and can be clearly assigned the class \emph{blue}: they lie within one sigma of the average energy when tested on the training dataset, which is shown as blue stripe around $E\approx-4$. Violet and green lie neither close to the blue or red region, so could be identified as ``neither''. Note that orange, yellow and warm reds have the highest energy and lie clearly within the identified ``red'' region around $E\approx3$. This is to be expected: the Hamiltonian classifier has no training data to discriminate between these colors, so it extrapolates---i.e.~yellow being green and red without blue, so the presence of red \emph{and} absence of blue amplify to this outcome.

A somewhat more convolved picture can be seen when we focus on the ground state of the trained Hamiltonian $\op H$.
We are interested in the overlap of the ground state $\gs(\op H)$ with the test data;
in \cref{fig:9color-result-3}, we plot the probabilities of a joint measurement of all the output qubits in \cref{fig:9color-graph}.
While there is a general trend towards larger overlap for blue colors (see the black line as a moving average), the data points themselves are very noisy, and there is a number of clearly-identified blue hues that would not be labeled as such.
Similarly, there are a number of false positives, i.e.\ red, orange or green colors which are within one sigma of the mean probability for the blue training data; this might also indicate that a more sophisticated regression should be used to interprete the test outcomes, and not only a linear one.
However, we also see that the blue data is delineated more or less sharply from all other data, especially as compared to the energy expectation values in \cref{fig:9color-result-1,fig:9color-result-2}.
This is as we predicted in \cref{sec:opt-energy}: expectation value optimization (\cref{def:ham-B}) is useful only as far as the data can be discriminated by a linear classifier, whereas ground spaces (\cref{def:ham-A}) can encode highly non-linear correlations.

\subsection{Benchmarking Annealing Parameters}\label{sec:e-benchmark-parameters}
\begin{figure}[t!]
	\centering
	\begin{tikzpicture}[
	x=2.2cm,y=1.3cm
	]
	\def\annParams{{5,10,20,30,40,50,75,100,125,150}}
	\draw (-.6,.6) [gray] +(.1,-.1) -- +(-.25,.4);
	\draw (-.6,.6) [black] +(.3,0) node[above] {$n_T=$} +(-.2,.2) node[right,rotate=270] {$R=$};
	
	\foreach \r in {0,...,9} {
		\pgfmathsetmacro{\resolutionN}{\annParams[\r]}
		\draw (-.7,-\r) node[left] {$\resolutionN$};
	}
	\foreach \t in {0,...,5} {
		\pgfmathsetmacro{\trotterN}{\annParams[\t]}
		\draw (\t,.6) node[above] {$\trotterN$};
		
		\foreach \r in {0,...,9} {
			\pgfmathsetmacro{\resolutionN}{\annParams[\r]}
			\draw (\t,-\r) node[] {\includegraphics[width=1.9cm]{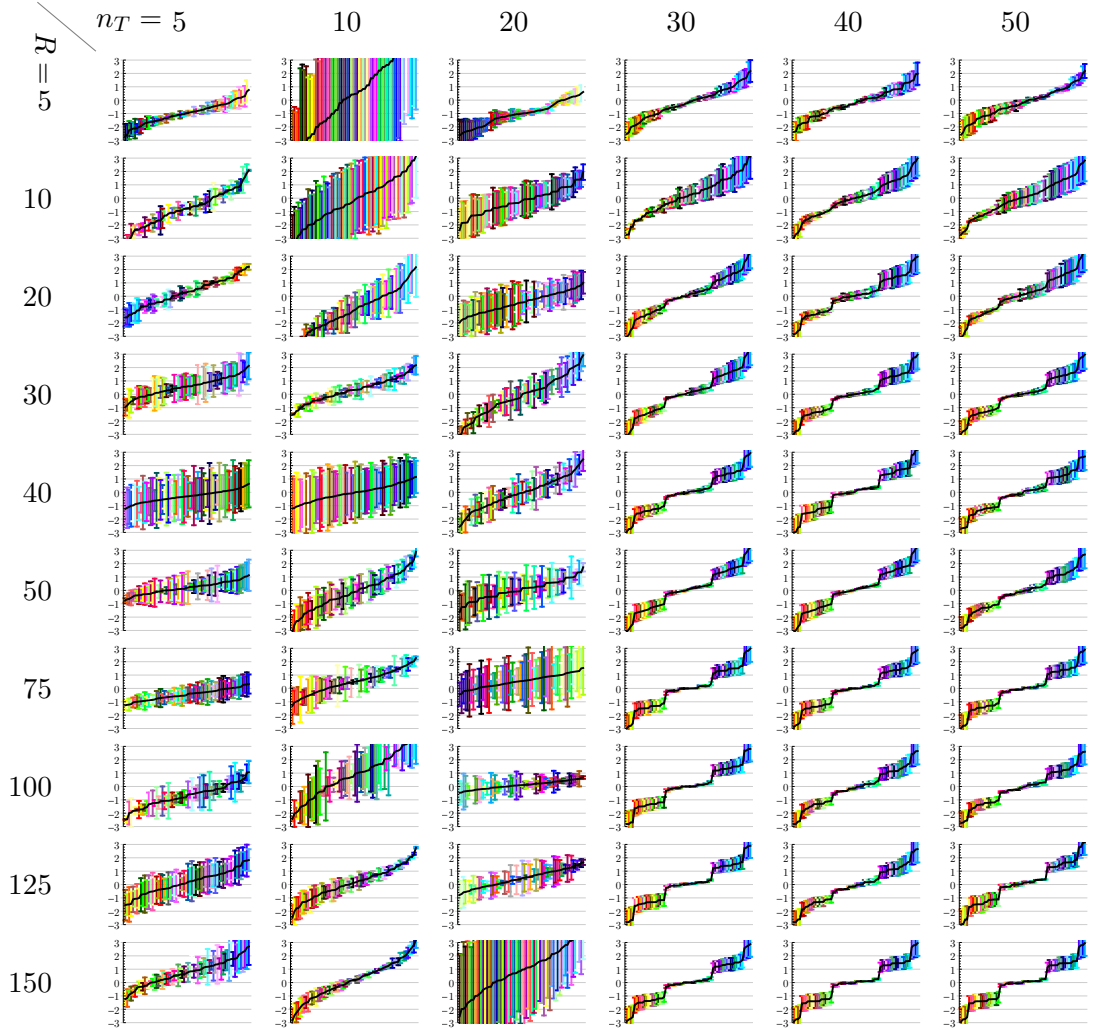}};
		}
	}
	\end{tikzpicture}
	\caption{6 colour classification performance for Trotter resolution $n_T$ and annealing step count $R$. We observe that for a trotter number $n_T\ge30$, there is essentially no improvement, and the resolution (i.e.\ annealing speed) plays a negligible role in this example.}
	\label{fig:annealing}
\end{figure}
The training scheme we proposed in \cref{sec:opt-groundstate} depends on fining the ground state of the training Hamiltonian, using (simulated) quantum annealing.
There are two meta-parameters:
\begin{enumerate}
	\item The annealing speed $R$, i.e.\ the velocity with which we tune the couplings from initial to target Hamiltonian in \cref{fig:annealing-schedule}.
	\item The Trotterization, i.e.\ the subdivision parameter $n_T$ in \cref{eq:trotter}.
\end{enumerate}
In order to test these parameters, we run a color classification task just as in \cref{sec:e-color} on six- instead of nine bit data.\footnote{This is solely done for performance reasons.}
Our findings are summarized in \cref{fig:annealing}.

One surprising insight is that the annealing speed seems to matter much less than the number of Trotter subdivisions $R$; as soon as $n_T\ge30$, the classification yields essentially the same results.

\subsection{Benchmarking Interactions}\label{sec:e-benchmark-interactions}
One interesting question which we have not addressed so far is what type of interactions and interaction topology is suited best for classification tasks.
So far, in \cref{sec:e-benchmark-parameters,sec:e-color}, we used diagonal projectors on custom-designed graphs (\cref{fig:9color-graph}).
But what if we use random interactions, or a Pauli basis?
In theory, if we allowed the interaction set in \cref{def:ham-A} to be a basis set
\[
    S_B=\{ \sigma_{e,1}\otimes\ldots\otimes\sigma_{e,|e|}: \sigma_{e,i}\in\{\1,\sigma^x,\sigma^y,\sigma^z\}, e\in E\},
\]
we would expect the Hamiltonian on that specific graph to be as optimal as possible; it is easy to prove that for any set $S'\neq S_B$,
\begin{align*}
	\sphericalangle(\gs(\op H(S')), \supp \Pi_\yes)       & - \sphericalangle(\gs(\op H(S')), \supp \Pi_\no)   \\
	\le  \sphericalangle(\gs(\op H(S_B)), \supp \Pi_\yes) & - \sphericalangle(\gs(\op H(S_B)), \supp \Pi_\no).
\end{align*}
However, in a practical implementation, it might not be feasible to tune all interactions at will;
e.g.\ in a D-Wave machine, the interactions are generally limited to Ising-type couplings.

\newcommand{\IG}[2][.6]{\includegraphics[scale=#1]{figures/#2}}
\newcommand{\IGG}[1]{\reflectbox{\rotatebox[origin=c]{180}{\includegraphics[width=3cm,height=.55cm]{figures/#1}}}}
\begin{table*}[t!]
	\centering
	\small
	\begin{tabular}{ llr | r S[table-number-alignment = left,table-figures-exponent = 1,table-figures-decimal=1] | lll}
		\toprule[1pt]
		\multirow{2}{*}{\textbf Graph}                 & \multirow{2}{*}{$\mathbf{S_e}$} & \multirow{2}{*}{\textbf \#} & \multicolumn{2}{c|}{\textbf Simulation} &         \multicolumn{3}{c}{\textbf Simulation Result}          \\
		                                           &                                 &                         & $N$ (opt) & {$t_\text{avg}$ [s]}    & Overlap               & {$f$ [\%]} & {$\Delta_\mathrm{E}$} \\ \midrule[1pt]
		\multirow{5}{*}[-4mm]{\IG{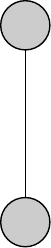}}              & \emph{Pauli}                    &                       3 & 18 (7)    & 360                     & \IGG{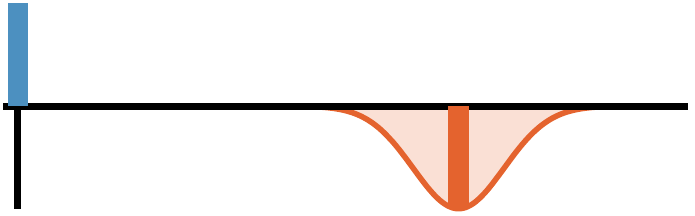} & 100        & 9.16                  \\
		                                           & \emph{Proj.}                    &                       3 & 6 (5)     & 11                      & \IGG{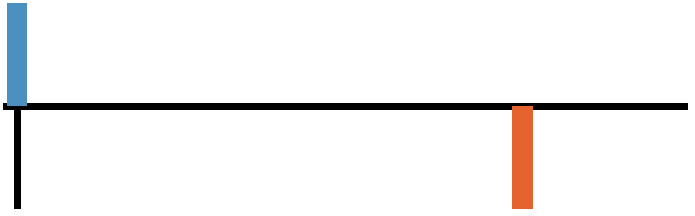}  & 100        & 2.0                   \\
		                                           & Rand.                           &                       3 & 9 (5)     & 11                      & \IGG{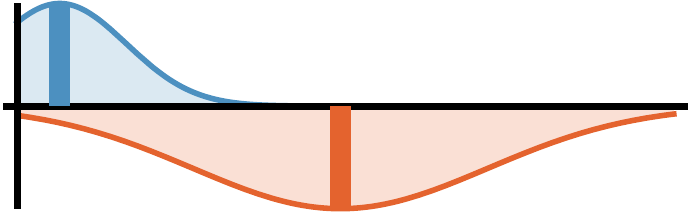}  & 88         & 1.95                  \\
		                                           & Heis.                           &                       3 & 5 (4)     & 1                       & \IGG{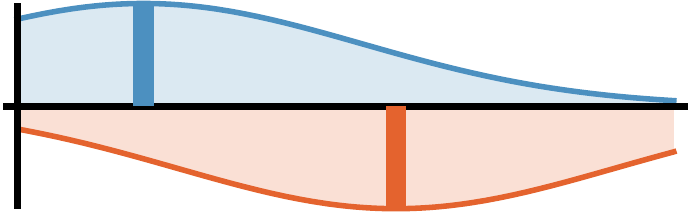}  & 75         & 1.37                  \\
		                                           & Ising                           &                       3 & 5 (4)     & 1                       & \IGG{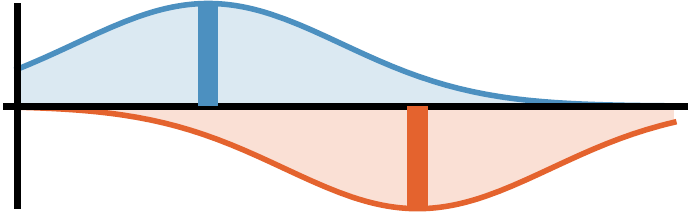} & 75         & 1.5                   \\ \midrule
		\multirow{5}{*}[-3mm]{\IG{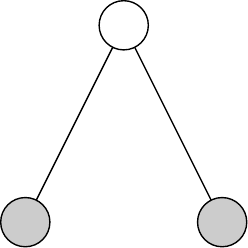}}              & Pauli                           &                       3 & 35 (13)   & 1.0e3                   & \IGG{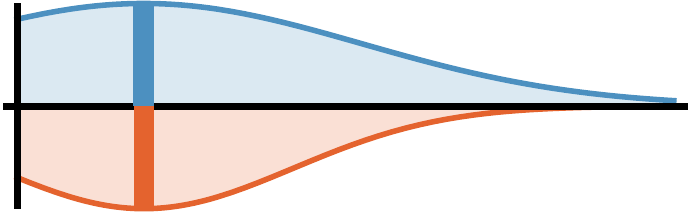} & 63         & 7.34                  \\
		                                           & Proj.                           &                       3 & 11 (7)    & 260                     & \IGG{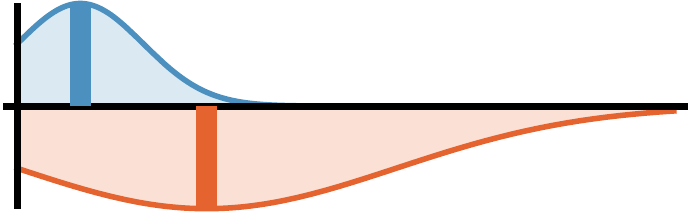}  & 63         & 1.49                  \\
		                                           & Rand.                           &                       3 & 17 (7)    & 310                     & \IGG{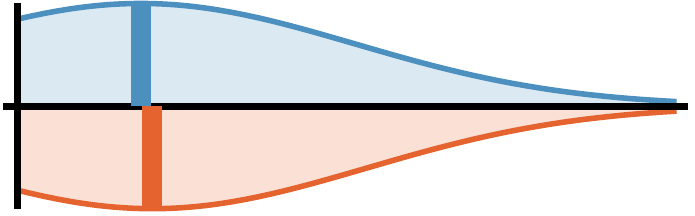}  & 50         & 3.37                  \\
		                                           & \emph{Heis.}                    &                       3 & 9 (6)     & 57                      & \IGG{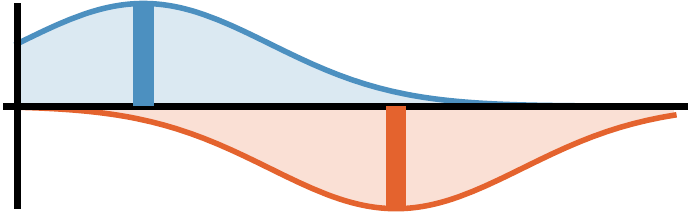}  & 75         & 0.0                   \\
		                                           & \emph{Ising}                    &                       3 & 8 (6)     & 57                      & \IGG{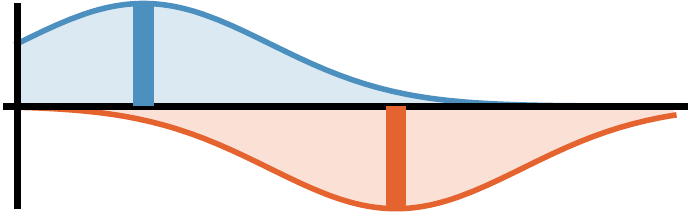} & 75         & 0.0                   \\ \midrule
		\multirow{5}{*}[-2mm]{\IG{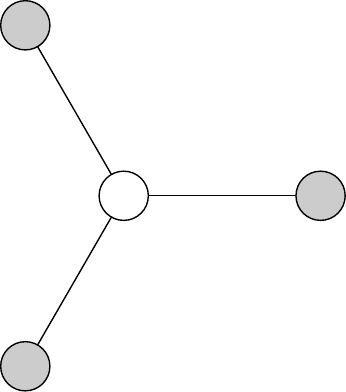}}              & \emph{Pauli}                    &                      35 & 52 (19)   & 15e3                    & \IGG{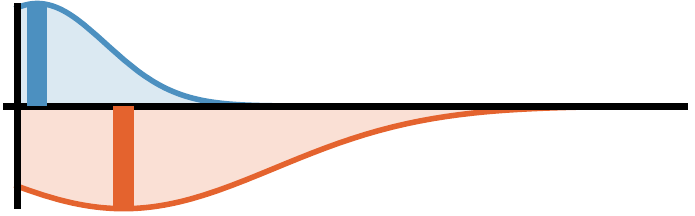} & 70         & 9.4                  \\
		                                           & Proj.                           &                      35 & 16 (8)    & 690                     & \IGG{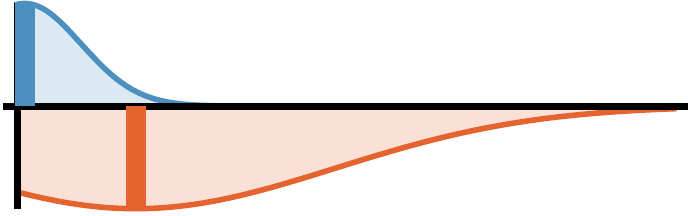}  & 63         & 2.2                   \\
		                                           & Rand.                           &                      35 & 25 (9)    & 8.7e3                   & \IGG{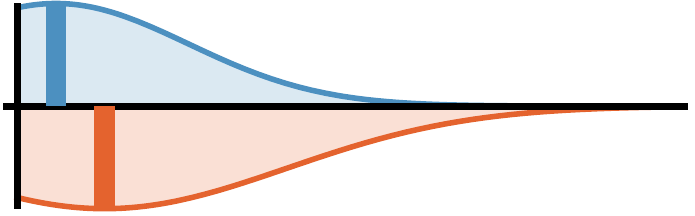}  & 57         & 3.68                  \\
		                                           & Heis.                           &                      35 & 13 (8)    & 840                     & \IGG{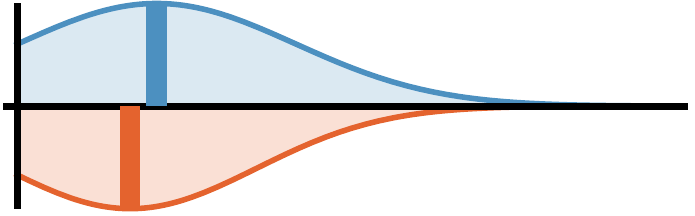}  & 54         & 0.0                   \\
		                                           & Ising                           &                      35 & 11 (7)    & 120                     & \IGG{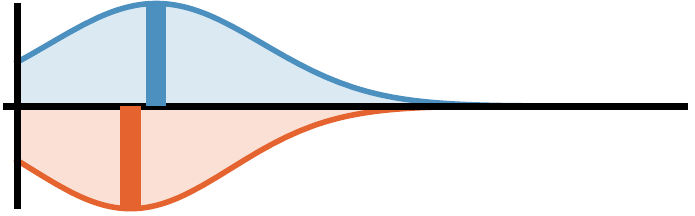} & 56         & 0.0                   \\ \midrule
		\multirow{5}{*}{\vspace{10mm}\IG{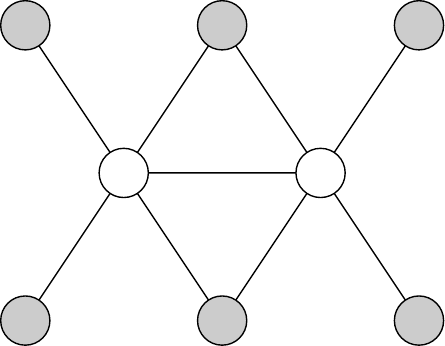}} & Proj.                           &                      10 & 44 (20)   & 35e3                    & \IGG{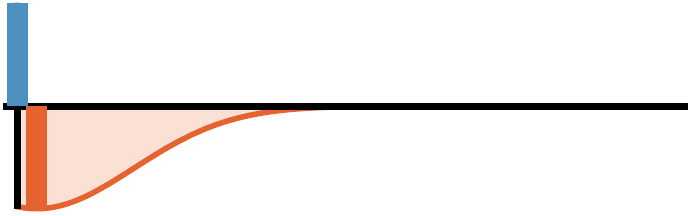}  & 54         & 2.11                  \\
		                                           & \emph{Rand.$^\dagger$}          &                       6 & 71 (22)   & 144e3                   & \IGG{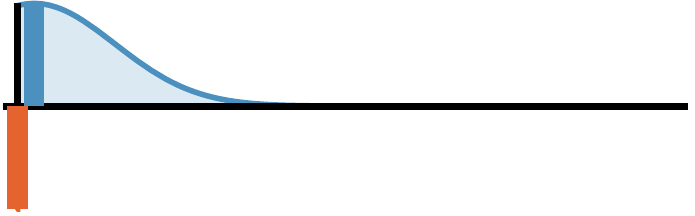}  & 69         & 4.46                  \\
		                                           & Heis.                           &                      10 & 35 (16)   & 11e3                    & \IGG{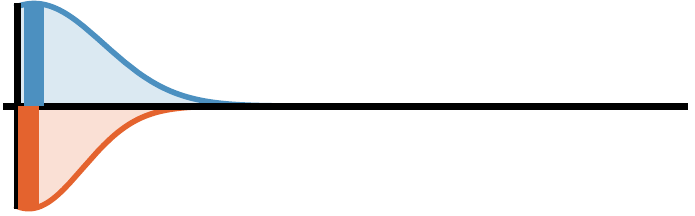}  & 57         & 0.33                  \\
		                                           & Ising                           &                      10 & 25 (14)   & 4.4e3                   & \IGG{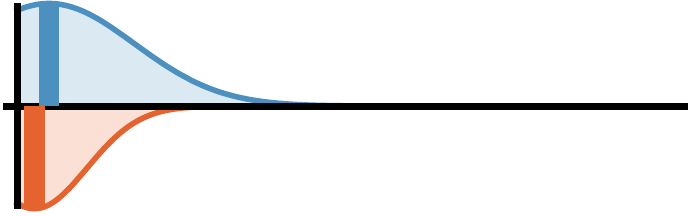} & 53         & 0.01                  \\ \bottomrule[1pt]
	\end{tabular}
	\caption{Interaction benchmark results.
		$S_e$ denotes the interaction set on the edges, which can be seen in \cref{eq:h1,eq:h2,eq:h3,eq:h4,eq:h5}. The best-performing interaction set for each graph is highlighted in italics.
		\# denotes the number of training data sets used, and $N$ the number of qubits in the training graph; (opt) gives the optimized graph qubit count, obtained by the qudit training scheme in \cref{sec:optimization}.
		The plot shows mean ground space overlap for \yes and \no data (orange/top and blue/bottom, respectively), as well as a fitted Gaussian showing the spread of the test results.
		To quantify the fidelity ($f$), we calculate the percentage of instances that can be classified correctly using a linear regression based on an expected ground space overlap we obtain from the training data, with the given graph and interaction type, where we average over all used training sets.
		Finally, $\Delta_\mathrm{E}$ is the average difference between the mean energy expectation of \yes and \no training sets with respect to the trained Hamiltonian.
		$^\dagger)$\ Due to lack of computation time, this test had to be aborted after six test sets.
		}
	\label{tab:benchmark}
\end{table*}

We therefore selected a series of interaction graphs small enough to perform our simulated annealing procedure, and tested whether any of the following two-local interaction sets outperformed the others.
\begin{align}
S_\text{Proj.} &:= \{ \ketbra{00}, \ketbra{01}, \ketbra{10}, \ketbra{11} \} \label{eq:h1} \\
S_\text{Pauli} &:= \{ \sigma_i\otimes\sigma j : i, j\in\{0, x, y, z\} \} \label{eq:h2} \\
S_\text{Rand.} &:= \{ \text{7 random Hermitian matrices on $\field C^4$} \} \label{eq:h3} \\
S_\text{Heis.} &:= \{ \sigma_i\otimes\sigma_i : i\in\{ x, y, z \} \} \label{eq:h4} \\
S_\text{Ising} &:= \{ \sigma_x\otimes\sigma_x, \sigma_z \} \label{eq:h5}
\end{align}
Our findings are summarized in \cref{tab:benchmark}.
One interesting observation is that there is no overall winner, i.e.\ the graph topology and the type of interaction seems to correlate.
Another point to note is that despite our claim that a full Pauli basis is always better than any other non-complete set of couplings, this seems to be violated here:
especially for the second graph, Pauli matrices perform much worse than Heisenberg or Ising-type couplings.
We reckon that this is due to the fact that our training algorithm is \emph{not} converging to a fixed point that optimizes \cref{def:ham-A}, especially for non-diagonal interactions---as we already suspected in \cref{sec:opt-groundstate}.

\subsection{Reducing Qubit Count for Small Devices}\label{sec:optimization}
The reader will have noticed by now that it sounds unfeasible to train even the diagonal projectors on the nine color graph in \cref{fig:9color-graph}.
If every one of the nine edges has to be optimized over projectors, then $|S_e|=4$ for all $e\in E$:
in addition to 10 spins of the graph we require $4\times 9=36$ control qubits, i.e.\ $46$ qubits overall.
This is clearly beyond the hard-coded limit of 23 qubits in \Liquid, and also well outside the realm of any supercomputer, in terms of memory requirement.

In order to still be able to train the graph weights, we employ the following approximation.
Assume one has $d$ interactions $\op h_1,\ldots,\op h_d$, acting on $\H$, that one wishes to obtain the coupling strengths for.
Our standard training Hamiltonian in \cref{eq:controlled-H} is
\[
	\op H_\text{train} = \sum_{i=1}^d \ketbra{1}_i\otimes \op h_{i} =
	\begin{pmatrix}
	0 & 0 \\ 0 & \op h_1
	\end{pmatrix}_{\!\!1}
	\otimes\ldots\otimes
	\begin{pmatrix}
	0 & 0 \\ 0 & \op h_d
	\end{pmatrix}_{\!\!d},
\]
where the matrix subscript labels the part of the Hilbert space $(\field C^2)^{\otimes d}\otimes\H$ the control acts non-trivially on.
For $d$ interactions to train, we need $d$ extra ancilla qubits.
An alternative approach is to write a control qu\emph{dit} instead of several control qu\emph{bits}, i.e.
\begin{equation}\label{eq:control-H-qudit}
	\op H_\text{train}' := \sum_{i=1}^d \ketbra{i+1} \otimes \op h_i =
	\diag\begin{pmatrix}
	0 & \op h_1 & \op h_2 & \cdots & \op h_d
	\end{pmatrix}.
\end{equation}
The control qudit has dimension $d+1$ and can thus be embedded in $\lceil \log_2(d+1) \rceil$ qubits, which is a significant reduction in the number of required ancillas.
This method can obviously be mixed with the full training Hamiltonian: if we e.g. group interactions in tuples of three (or seven), we can train them with two (or three) control qubits, each.

One immediate caveat is that in contrast to \cref{eq:controlled-H}, \cref{eq:control-H-qudit} is  not a sum over all possible combinations of interactions; for instance, there would never be a block in the overall training matrix containing a combination of $\op h_3+\op h_d$.
We did not find this to be a problem when training our nine color example, but want to point out that we could only test this case with diagonal projectors---which could indeed be a special case.
In \cref{sec:e-benchmark-interactions}, we give a series of smaller graphs which we benchmark with this qudit optimization scheme and more general interaction sets, many instances of which we note could only be simulated due to the improved qubit counts in contrast to the vanilla technique.

\section{Discussion and Outlook}
In this work, we have proposed and discussed two data classification models: ground state overlap and energy expectation with respect to a family of local Hamiltonians, where we optimize over its coupling constants.
We have shown that the model with energy expectation value is computationally easy and thus unlikely to capture much computational power beyond a linear program (\cref{sec:opt-energy}).
In contrast, we have motivated and empirically demonstrated that the model where ground state overlap is maximized for \yes instances in the dataset, and minimized for \no instances, seems to bear much more inherent complexity, making it suitable for discriminating data on a more sophisticated level.
We verify this claim empirically;
due to the lack of sufficient computational power or access to an actual quantum annealing device, we had to restrict our analysis to small toy problems.
In particular, we hope that for future iterations we can test the sequential learning algorithm in \cref{sec:opt-groundstate}, which we expect to outperform our current training algorithm, and perform some tests on a noisy quantum device.

We would like to emphasize that the number of ancillas necessary for our training scheme only grows at most linearly with the number of interactions in the model---our proposed method for obtaining Hamiltonian coupling constants is thus still comparatively cheap, despite the difficulty of simulating the annealing procedure on classical hardware.
One caveat is that it seems unlikely that the scheme---as it currently is---is optimal for non-diagonal interactions, and more work has to be done to generalize it to other sets of couplings.
In the same spirit, a theoretical result showing (approximate) convergence towards terms that optimize \cref{def:ham-A} is still missing.

In a sense, the question of finding a ground state overlap-maximizing Hamiltonian is related to finding a parent Hamiltonian to a given set of states (i.e.\ finding a Hamiltonian $\op H$ with $\op H\ket\psi=0$ for all $\ket\psi$ in some set), an ongoing field of research---e.g.\ with results on matrix product states \cite{Perez-Garcia2007,Fernandez-Gonzalez2015}, quantum error correction \cite{Brandao2017}, or lattice crystals \cite{Huerga2017}.
While related, we impose the additional constraint that we are given a fixed set of local interactions which we optimize, and that we simultaneously want to \emph{minimize} overlap with another set of wave functions (the \no data);
nevertheless, we hope that new insights from the research on parent Hamiltonians will help us to further develop our ideas.

Last but not least, we want to present the reader with three open questions.
\begin{enumerate}
\item How can we modify our proposed training scheme in \cref{sec:empirical} to yield better results for non-diagonal interactions?
Can we find an analytical expression for its adiabatic convergence?
\item As soon as we allow for an interaction set which includes two-or-more local couplings, the training Hamiltonian \cref{eq:controlled-H} is at least three-local.
However, in the circuit model, we know how to break down unitaries acting on $d\ge3$ qubits into unitaries of locality at most $2$ (e.g.\ using Solovay-Kitaev).
Can we adapt our training Hamiltonian in a similar fashion?
Will we need methods such as perturbation gadgets in this case (see e.g.\ \cite{Piddock2015})?
\item Can we train this model on an actual piece of quantum hardware, such as a D-Wave machine?
What data can we classify, and how does this compare to a similarly-sized (e.g.\ in the number of parameters) classical algorithm?
\end{enumerate}

One interesting hybrid approach would be to see whether one can obtain a better classifier when looking at thermal states of the Hamiltonian.
As we have seen in \cref{sec:empirical}, the ground state overlap with test data is very noisy (\cref{fig:9color-result-3}), whereas the energy expectation is not (\cref{fig:9color-result-1}).
This might indicate that if our training scheme fails to push a data point to the ground state, it is still close, in the sense of being in the high weight area of a Gibbs state with appropriately-chosen temperature.

As common with conceptual propositions, there are more new questions that arise than old ones that can be answered.
We realize that most of our contribution is of empirical nature, in the spirit of ``experimental'' computer science research.
However, given the success of the latter, we hope that our work is inspiration for further thought on these topics.

\section*{Acknowledgments}
J.\,B.\ acknowledges support from the German National Academic Foundation, the EPSRC (grant 1600123), and the Draper's Research Fellowship at Pembroke College.
We would also like to thank the organizers of the Microsoft \Liquid competition 2016 for inspiring this work, the CWI in Amsterdam and mathsQI in Madrid for inviting a talk on this topic, and for the invaluable feedback we gained from discussions there.

\printbibliography

\end{document}